\documentclass[fleqn,usenatbib]{mnras}

\usepackage{newtxtext,newtxmath}
\usepackage[T1]{fontenc}
\usepackage{ae,aecompl}

\usepackage{graphicx}	% Including figure files
\usepackage{xcolor}
\usepackage{ulem}

\newcommand\msun{M_{\odot}}
\newcommand\rsun{R_{\odot}}
\newcommand\LISA{\textit{LISA}}
\newcommand\Zsun{Z_{\odot}}
\newcommand\fGW{f$_{\mathrm{GW}}$}
\newcommand{\SNR}{S/N}
\title[DWDs in the MW]{Predicting the LISA white dwarf binary population in the Milky Way with cosmological simulations}%{Predicting white dwarf binaries detected by \textit{LISA} with cosmological simulations}

\author[A. Lamberts et al.]{
Astrid Lamberts$^{1,2}$\thanks{E-mail: astrid.lamberts@oca.eu},
Sarah Blunt$^{3,4,5}$,
Tyson B. Littenberg$^{6}$,
Shea Garrison-Kimmel$^{2}$,
\newauthor Thomas Kupfer$^{7}$,
Robyn E. Sanderson$^{8,9}$
\\
$^{1}$ Universit\'e C\^ote d'Azur, Observatoire de la C\^ote d'Azur, CNRS, Laboratoire Lagrange, Laboratoire Art\'emis, France\\
$^{2}$ Theoretical Astrophysics (TAPIR), California Institute of Technology, Pasadena, CA 91125, USA\\
$^{3}$ NSF Graduate Research Fellow \\
$^{4}$ California Institute of Technology, Pasadena, CA 91125, USA\\
$^{5}$ Center for Astrophysics | Harvard\& Smithsonian, Cambridge, MA 02138, USA\\
$^{6}$ NASA Marshall Space Flight Center, Huntsville, AL 35812, USA \\
$^{7}$ Kavli Institute for Theoretical Physics, University of California, Santa Barbara, CA 93106, USA\\
$^{8}$ Department of Physics \& Astronomy, University of Pennsylvania, 209 S 33rd St., Philadelphia, PA 19104, USA\\
$^{9}$ Center for Computational Astrophysics, Flatiron Institute, 162 5th Ave, New York, NY 10010, USA
\\
}

\date{Accepted XXX. Received YYY; in original form ZZZ}

\pubyear{2019}

\begin{document}
\label{firstpage}
\pagerange{\pageref{firstpage}--\pageref{lastpage}}
\maketitle

\begin{abstract}

Short-period (P$_{\mathrm{orb}}$<1 hour) white dwarf binaries will be the most numerous sources for the space-based gravitational wave detector~\LISA. Based on thousands of  resolved systems, we will be able to constrain binary evolution  and provide a new map of the Milky Way and its surroundings. Here we predict the main properties of populations of different types of detached white dwarf binaries detected by~\LISA. For the first time, we combine a high-resolution
cosmological simulation of a Milky Way-mass galaxy (from the FIRE project)
with a binary population synthesis model for low and intermediate mass stars. Our model therefore provides a cosmologically realistic star formation and metallicity history for the galaxy and naturally produces its different components such as the thin and thick disk, the bulge, the stellar halo, and satellite galaxies and streams. With the simulation, we show how different galactic components contribute differently to the gravitational wave signal,  due to their typical age and distance distributions. We find that the dominant~\LISA~sources will be He-He systems and He-CO systems with important contributions from the thick disk and bulge but also a few systems in the stellar halo. The resulting sky map of the sources is different from previous models, with important consequences for the searches for electromagnetic counterparts and data analysis. We also emphasize that much of the science-enabling information regarding white dwarf binaries, such as the chirp mass and the sky localisation, becomes increasingly rich with long observations, including an extended mission up to 8 years.

\end{abstract}


%%%%%%%%%%%%%%%%%%%%%%%%%%%%%%%%%%%%%%%%%%%%%%%%%%

\section{Introduction}
\label{sec:intro}

Gravitational waves (GW) are the most promising way towards systematic detection of compact binaries. The LIGO/Virgo detectors have observed the mergers of several binary black holes \citep{LVC_19_GWTC1} and a binary neutron star \citep{LIGO_BNS_170817}, emitting GW in the kiloHertz regime. Within the next 20 years the \textit{Laser Interferometer Space Antenna (LISA) } will open up a new window in the GW spectrum, between 10$^{-5}$ and 10$^{-2}$ Hz \citep{LISA_proposal}.  By numbers, the dominant sources for \textit{LISA} will be double white dwarfs (DWD) in our Milky Way (MW), about a hundred thousand years before they merge. As white dwarfs (WD) are the remnants of stars below $\lesssim 8\msun$, more than 95$\%$ of the stars are likely to end their lives as WDs.  

In a seminal paper, \citet{nelemans01_WD} determined several tens of millions of detached DWDs would be present in the \textit{LISA} band and roughly ten thousand of them, with GW frequency \fGW$\gtrsim 0.4$ mHz, would be individually resolvable. With (at least) thousands of detectable systems, \textit{LISA} will allow new statistical studies of close DWDs. Such studies will strongly advance our understanding of stellar and binary evolution. The distribution of chirp masses and periods will allow to constrain the impact of the common envelope, which drastically tightens the orbit of the systems \citep{Toonen:2014aa}. A complete sample will also allow for a direct comparison with the post-common envelope binaries, which have undergone only one episode of the common envelope \citep{Rebassa_mansergas_12_PCEbinaries}. In some cases, the frequency derivative will be measurable and will allow to determine if mass transfer is happening \citep{Breivik_2018_accretion} and/or tidal interactions are deforming the white dwarfs.

GW observations are very complementary to electromagnetic (EM) observations, which are challenging as WDs are faint and rapidly cool down to become even fainter. Even with dedicated surveys, our view of DWDs in the MW is going to be hindered by dust extinction and faintness of the sources before the start of \textit{LISA} operations. Short period binaries observable by~\LISA~(orbital period below half an hour) are found with phase-resolved spectroscopy of previously discovered white dwarfs \citep{Napiwotzki_01_SPY,Brown_10_ELM1,Brown_16_ELM7} or light curves from high cadence surveys \citep{Levitan_13_PTD_DWD}.  Roughly 20 DWDs have been discovered with a high enough frequency to be detectable by \textit{LISA}. Most of these binaries are interacting binaries, which are a rare sub-class but are easier to detect electromagnetically because of the presence of an accretion disk \citep{nelemans01_WD}. 
These electromagnetically identified binaries are called "verification binaries" and are guarunteed multimessenger sources~\citep[see][for an updated list using Gaia distances]{Kupfer_18_Gaia_LISA}. Large scale systematic searches for these high frequency systems are just starting, withe e.g. the high cadence survey ZTF (Zwicky Transient Factory; \citealt{bellm:2019,graham:20119}) and possibly LSST (Large Synoptic Survey Telescope).

DWDs will be a new way to look at our MW, showing a population of  older, low mass stars. As the strain amplitude of GW  decreases only as $1/r$ (in comparison to $1/r^2$ decrease for electromagnetic emission),~\LISA~will be able to more easily sample more remote regions of our Galaxy, its satellite and maybe Andromeda \citep{Cooray_05_LISA_LMC,Korol_18_LISA_LocalGroup}. The~\LISA~ detections could lead to a new measurement of the Galactic potential \citep{Korol_18_MWpotential} and the global amplitude of the signal due to DWDs will quantify the star formation history of the MW~\citep{Yu_2013_SFH_DWD}.

Aside from their importance for stellar/binary evolution and the Galactic structure, predicting and understanding the GW detections of DWDs is crucial to the success of the~\LISA~mission. Most of the DWDs will be unresolved, meaning there will be more than one binary emitting in a given frequency bin, which width is set by the inverse of the observation time (roughly the duration of the mission). Below $\simeq 2$ mHz, the combination of these unresolved sources will effectively be a contaminating foreground which will prevent or hinder the detection of other sources such as extreme mass ratio inspirals or supermassive black hole mergers at low masses \citep{nelemans01_WD,Marsh_11_DWD_LISA,Ruiter_2010_DWD,Nissanke_2012_DWD}.

Since the first predictions based on a Galaxy model combined with a binary population model~\citep{nelemans01_WD,nelemans_01_AMCVn}, models have included detailed studies of different DWD formation channels~\citep{Nissanke_2012_DWD}, the different types of DWDs and their spatial distribution in the MW~\citep{Ruiter_2010_DWD}. Important uncertainties remain regarding binary evolution \citep{Postnov14_review}, although the volume of observational completeness in our neighbourhood is slowly increasing and is a promising way to put constraints~\citep{Toonen_17_localDWD}. More recent studies demonstrate the potential of multimessenger detections and the link with \textit{Gaia} and LSST \citep{Korol17_WDLISA,Breivik_2018_accretion}. \citet{Korol_19_disk_bulge} predicts that \textit{Gaia} will detect about 25 verification binaries within 2 kpc, and LSST about 50 more, within 10 kpc; and that most of them will be away from the Galactic plane and bulge.

All these studies are based on parametrised models for the MW's star formation and structure. They use axisymmetric models for the different components of the galaxy, which often only model the thin disk and bulge. Star formation usually follows the~\citet{Prantzos_2000_MW_SFR} star formation model of the MW and assumes a unique value of the metallicity for each galactic component. \citep{Ruiter_09_halo} first highlighted that different galactic components have different contributions to the GW signal because of their different age, metallicity and typical distances.  These findings motivate the present analysis, where we combine a binary population synthesis model with a cosmological hydrodynamic simulation of a MW-like galaxy \citep{Wetzel2016} to model the structure and star formation history. This allows us to naturally include all the components of the MW such as the thin and thick disk, the bulge and the accreted stellar halo, as well as a population of satellite galaxies. A similar approach for binary black holes \citep{Lamberts_18_BBH_MW} has shown that the latter are over-represented in the stellar halo of the galaxy, where the metallicity is low. 

This paper builds on the methodology developed in \citet{Lamberts_18_BBH_MW} combining synthetic binary black hole populations with the same cosmological simulations (\S\ref{sec:method}). We will show the resulting detached DWD populations and how their main properties stem from binary evolution and galactic structure and evolution (\S\ref{sec:results}). We will highlight the impact of a complete Galactic model for the detection of GW with~\LISA~(\S\ref{sec:GW}) and  compare it with previous results (\S\ref{sec:discussion}) and conclude (\S\ref{sec:conclusions}).

\section{Method}
\label{sec:method}
We follow the same method as \citet{Lamberts_18_BBH_MW}, built on  a set of simulations of MW-like galaxies (\S\ref{subsec:galaxy}) and a binary population synthesis model (\S\ref{subsec:BPS}) uniquely combined together (\S\ref{subsec:combination}) to make GW predictions (\S\ref{subsec:GW_methods}).

\subsection{FIRE Galaxy Model}
\label{subsec:galaxy}
We use a subset of MW-like galaxies from the Feedback in Realistic Environment (FIRE; \citealp{Hopkins:2014_FIRE}) project\footnote{\url{http://fire.northwestern.edu}}, including \textbf{m12i} (\emph{a.k.a.} the ``Latte'' simulation; \citealp{Wetzel2016}, \textbf{m12m} and \textbf{m12f} \citep{GarrisonK_17_embedded_disk} simulations.
These simulations are based on the  improved ``FIRE-2'' version of the code from \citet[][for details, see Section~2 therein]{Hopkins2017fire2} and ran with the code GIZMO \citep{Hopkins2015gizmo}\footnote{\url{http://www.tapir.caltech.edu/~phopkins/Site/GIZMO.html}}. GIZMO solves the equations of  hydrodynamics using the mesh-free Lagrangian Godunov ``MFM'' method.  The analysis of the simulations is done with the publicly available Python package \texttt{gizmo\_analysis}.

More specifically, these simulations have an initial gas particle mass of about 7070 M$_{\odot}$ and for the gas, both the hydrodynamic and gravitational (force softening) resolutions are fully adaptive down to 1 pc. The simulations include cooling and heating from a meta-galactic background and local stellar sources from $T\sim10-10^{10}\,$K.  Star formation occurs in locally self-gravitating, dense, self-shielding molecular, Jeans-unstable gas. Stellar feedback from OB and AGB star mass-loss, type Ia and II supernovae, and multi-wavelength photo-heating and radiation pressure is  directly based on stellar evolution models. Chemical enrichment stems from type Ia supernovae \citep{Iwamoto99_SNyields}, core-collapse supernovae
\citep{Nomoto06_SNyields}, and O and AGB star winds \citep{VandenHoek99_AGByields,Marigo01_AGByield,Izzard04_AGByield}. The simulations include subgrid-scale numerical turbulent metal diffusion terms \citep{Hopkins2017fire2,Bonaca2017}, which have almost no dynamical effect at the galaxy mass scales considered here \citep{Su2017_metaldiff}, but produce better agreement with the internal metallicity distribution functions observed in MW satellite galaxies \citep{Escala2017_dmdf}. All the binary evolution models are included during post-processing, and the hydrodynamic simulation does not explicitly include binary effects. 
 
Our analysis is based on galaxy {\bf m12i} (from \citealp{Wetzel2016}), though we analyze a re-simulation with turbulent metal diffusion first presented in \citealp{Bonaca2017}, chosen to have a merger history comparable to the Milky Way. We also consider a lower-resolution version of {\bf m12i} as well as two different galaxies {\bf m12f} and {\bf m12m} \citep{Hopkins2017fire2} at the same mass scale. {\bf m12i} shows metallicity gradients \citep{Ma17_metalMW} and abundances of $\alpha$-elements (Wetzel et al, in prep.) in the disk that are broadly consistent with observations of the MW. Its global star formation history is consistent with the MW (see \citealp{Ma17_metalMW} for illustrations) although its present day star formation rate of 6$\msun$ yr$^{-1}$ is somewhat higher than observed in the Milky Way. The satellite distribution around the main galaxy in {\bf m12i} presents a similar mass and velocity distribution as observed around the Milky Way and M31, down to a stellar mass of $10^5\msun$, though the simulation does not contain an equivalent of the Large Magellanic Cloud; the most massive satellite is comparable to the Small Magellanic Cloud. Outputs from the simulations and corresponding mock Gaia catalogs are available online \footnote{https://fire.northwestern.edu/data/ and  http://ananke.hub.yt}, based on  \citet{Sanderson2018}, which also compares the simulated galaxies with the Milky Way. Effectively our analysis is based on this publicly available data, except for the information on the location of the stars at their formation, which have been obtained with permission.

From the simulation, we recover the position, formation time $t_*$, metallicity $Z$ and position and mass at formation $M_*$ of every star particle\footnote{Whenever we refer to the simulation, we use the words star, particle and star particle interchangeably.}. We only use the particles within 300 kpc of the center of the galaxy. This is slightly larger than the virial radius of the galaxy and allows us to largely sample the halo, satellites and streams while remaining unaffected by the boundaries of the high resolution region. This yields a list of roughly 14 million star particles.

The simulations assume a $\Lambda$CDM cosmology with $\Omega_{\Lambda}$ = 0.728, $\Omega_m$ = 0.272, $\Omega_b$ = 0.0455, h = 0.702, $\sigma_8$ = 0.807, and n$_s$ = 0.961 \citep{Planck16_cosmo}. All metallicities are defined with respect to the solar metallicity, set to $\Zsun=0.02$.

\subsection{Binary Population Synthesis model (BPS)} 
\label{subsec:BPS}

To simulate a population of DWDs, we use a modified version of the publicly available \texttt{BINARY STAR EVOLUTION (BSE)} code based on the rapid binary evolution algorithm described in \citet{Hurley:2002aa}. We only consider formation through binary evolution as the survival of compact low mass binaries is unlikely in dense stellar environments. For low mass binaries \citep[see][for a recent review]{Postnov14_review}, the main uncertainty stems from our limited understanding of the common envelope phase \citep{Ivanova:2013aa}.  

As in \citet{Lamberts_18_BBH_MW} we model 13 logarithmically spaced metallicity bins between $5\times 10^{-3}$ and 1.6$\Zsun$.  We model a distribution with a thermal eccentricity~\citep{Heggie_75_thermal_e}, which favors systems with high eccentricity and model a  distribution of initial separations between  1 $\rsun$ and 10$^6\rsun$ following a flat distribution in log space~\citep{Abt_83_periodsbinaries}. Primary masses $m_{1*}$ are drawn from a Kroupa IMF \citep{Kroupa:2001} between  $0.95<m_{1*}<10 \msun$ and secondary masses are set by $m_2=q m_1$ where $q$ is uniformly distributed between 0 and 1. We discard binaries with  $m_{2*}<0.5\msun$ as lower mass secondaries will not form DWDs within a Hubble time. With this condition, more than 90 per cent of the systems are discarded, saving significant computing time. We keep track of the number $\bar{N}_b$ and mass $\bar{M}_b $ of discarded binaries in order to normalize the number of DWDs to the stellar mass in our galaxy simulation (see \S\ref{subsec:combination}). 

We perform the population synthesis on  $N_b=$ 2.5 million systems per metalliticy bin. We performed convergence tests on the period distribution of DWDs at their formation and determined that 2.5 million binaries \textit{within the narrow mass range for the primary and secondary where DWD formation is possible}, is necessary to appropriately sample the tightest orbital periods. The latter have the highest GW frequency and will be very loud sources for~\LISA. 

We use \texttt{BSE} to evolve our population forward up to the current age of the universe,  tracking the systems that form binary white dwarfs.  We use the following binary evolution parameters and characteristics for our population synthesis:
\begin{itemize}
\item Tidal circularization is enabled.
\item  The Helium star mass loss factor is 1.
\item Mass loss for more massive stars is set by \citet{Vink:2001}, with the wind velocity factor $\beta=-1$.
\item We follow the common envelope evolution description from \citet{Tout_97_common_env} with efficiency parameter $\alpha$ set to 1 (see \citealt{Ivanova:2013aa}). The critical mass ratio to start a common envelope interaction is set by the polytrope solution by \citet{Tout_97_common_env} depending on the mass and radius of the star, and set to 0.25 for stars in the Hertzprung gap.
\item We assume Roche lobe overflow mass transfer is conservative.
\item Accretion onto a compact object has an efficiency of 0.5.
\item We set the Eddington limit for mass transfer to 1.
\item The mass of the WD at formation is naturally set by the competition between core-mass growth and envelope mass loss.
\end{itemize}
There are many important uncertainties, especially for the impact of mass transfer. In this work, we have chosen standard values for the binary and stellar evolution. As our focus is the combination with an updated model for the Milky Way rather than binary evolution, we restrict ourselves to this single set of parameters and leave a wider exploration for further work.

For each metallicity, we eventually produce a list of DWDs with their formation time after the formation of the progenitor binary, their orbital properties and masses. With 2.5 million initial binaries in a the appropriate mass range, we end up with about 700 000 DWDs in each metallicity bin. For a binary fraction of 0.5 we find a DWD formation rate of 0.012-0.016 DWDs per unit Solar mass of total star formation (including binaries and singles). There is limited variation with metallicity. 

We identify He (helium) WDs, CO (carbon/oxygen) WDs and Ne (neon) WDs separately. These different populations stem from different progenitor masses and/or binary evolution channels. Different subtypes of WDs have different radii and cooling times, which is important for their electromagnetic properties. In this paper we will show that different subtypes also contribute differently to the GW signal.

\begin{figure*}
	\includegraphics[width=.9\textwidth]{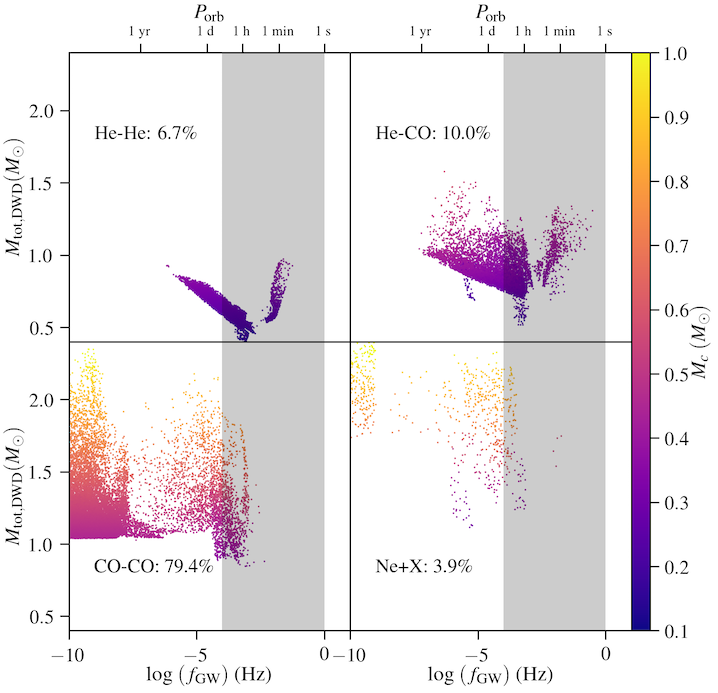}
    \caption{Final periods and masses for a binary evolution model with 2.5 million binaries for $Z=\Zsun$. Each quadrant shows a different subtype of resulting DWDs and the fraction of systems formed. We show the gravitational wave frequency $f_{\mathrm{GW}}$ and orbital period $P_{\mathrm{orb}}$ at the formation of the binary and its total mass $M_{\mathrm{tot,DWD}}$. The color shows the chirp mass $M_c$ which is relevant for detectability with \textit{LISA}. The gray vertical band shows the frequency range where~\LISA~will be sensitive.}
    \label{fig:final_Mtot_v_P}
\end{figure*}

\begin{figure*}
	\includegraphics[width=.3\textwidth]{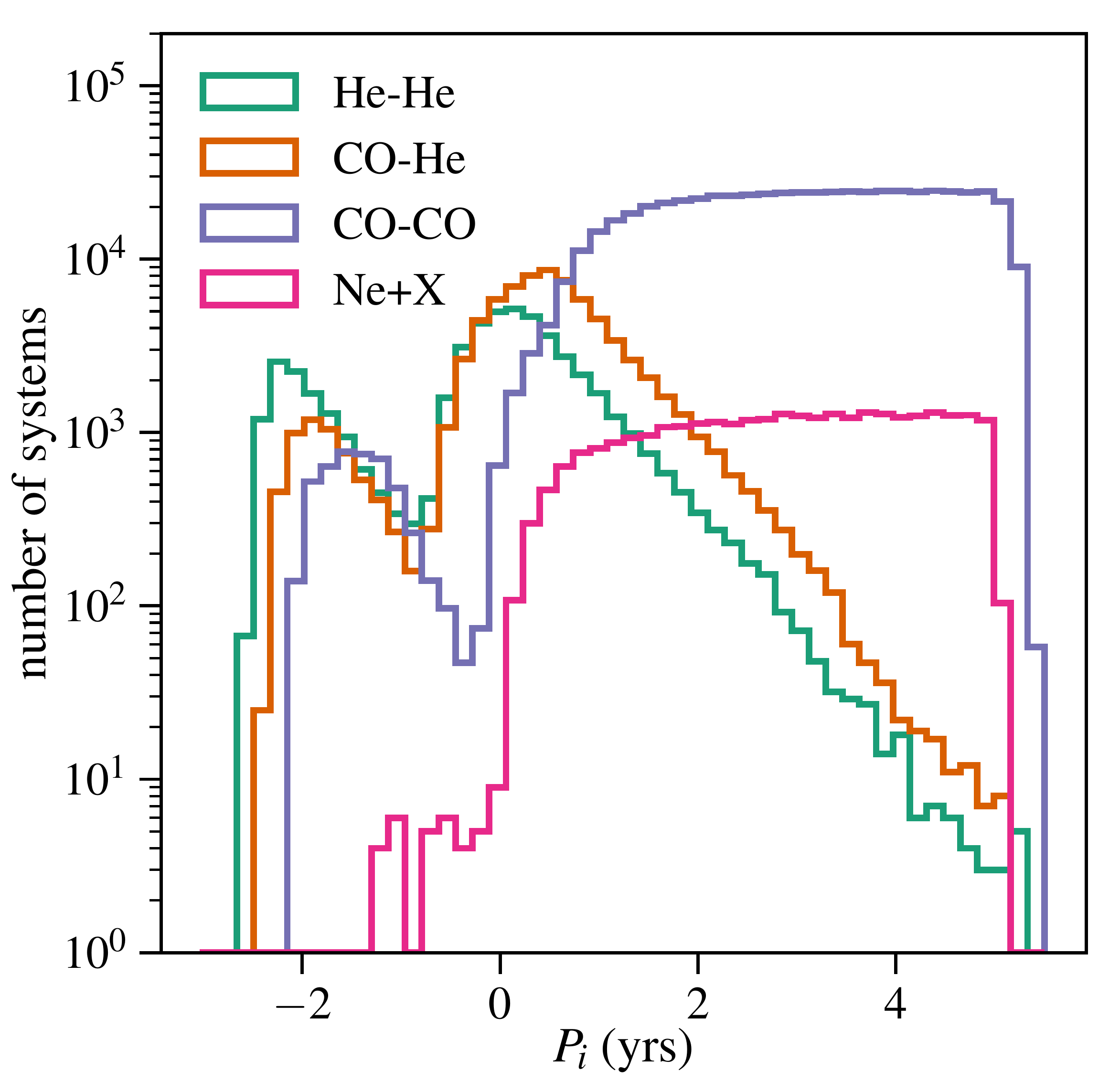}
	\includegraphics[width=.3\textwidth]{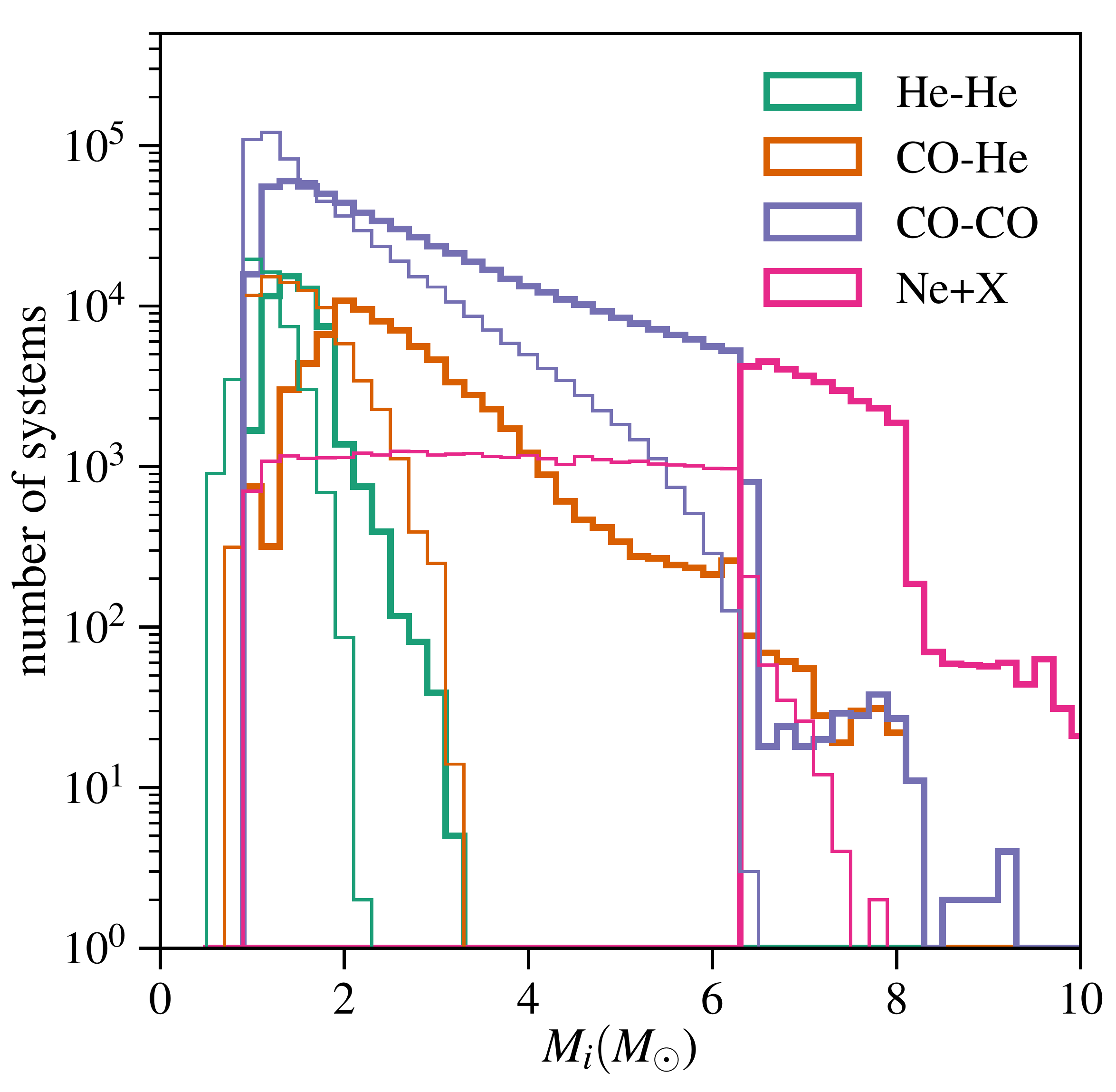}
	\includegraphics[width=.3\textwidth]{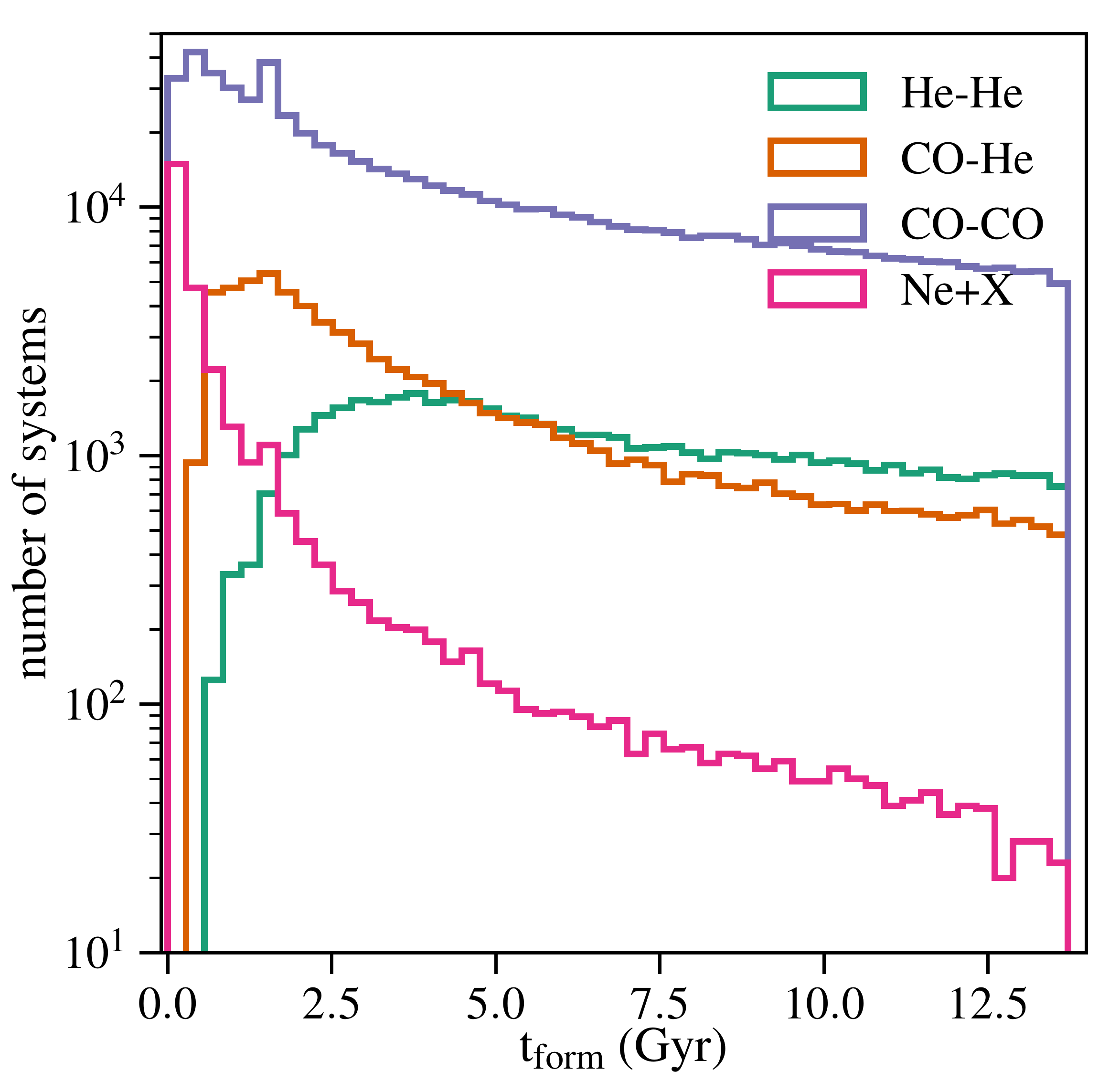}
    \caption{Initial period (left) and progenitor masses (middle) of the final DWDs as a function of the DWD subtype for $Z=\Zsun$. We show the primary mass (thick line) and secondary mass (thin line). The right panel shows the time needed to form the DWD.}
    \label{fig:prog_props}
\end{figure*}

Fig.~\ref{fig:final_Mtot_v_P} shows the masses and orbital periods at the formation of the DWD binary as computed by \texttt{BSE} for an initial population at Solar metallicity. In comparison, Fig.~\ref{fig:prog_props} shows the properties of the corresponding progenitor binaries. The first column in Tab.~\ref{tab:summary_numbers} shows the absolute numbers of DWDs created in \texttt{BSE}. We distinguish 4 types of binaries depending on the nature of its white dwarf components:
\begin{itemize}
    \item He-He WDs: These come from two low mass stars, which evolve very slowly, and have both their envelopes stripped by common envelope interactions. He-He DWDs stem from binaries with short initial periods (or high eccentricities allowing for short periastron passages) and constitute a small fraction of the total population of DWDs, but they are important for~\LISA. The formation time of these binaries is rather constant between 2 and 13 Gyrs, which is much longer than the other channels. This results in low mass WDs (M$_{\mathrm{WD}}<0.45$ $\msun$) in a very tight orbit. The tightest binaries are going to merge quickly due to their GW emission. They are also  going to interact tidally during later phases, because of their comparatively large radius.
    \item CO-CO WDs: These systems come from initially wider orbits, preventing the stripping of the envelope before the beginning of core He burning, and resulting in CO cores. Most of these systems have never interacted and will always have a large separation, which makes them less relevant for~\LISA. CO-CO binaries form in less than a Gyr and make up the bulk of the DWD population, with masses above $0.45\msun$ (and often above $0.65\msun$) and periods down to one hour. 
    \item He-CO WDs: These systems are a mixture of both previous categories. They need about 2 Gyrs to form and have low chirp masses because of their unequal masses. They can also form very tight DWDs and, combined with the He-He WDs, they are the most numerous in the~\LISA~band although they make up only 10 per cent of the global DWD population.
    \item Ne WDs: These are systems with at least one Ne/O WD, meaning that one of the stars has started carbon burning in the core before turning into a WD. As such, these WDs stem from massive stars: they form on short timescales, come from initially well-separated stars (to prevent stellar mergers) and are rare.
\end{itemize}

\begin{table*}
	\centering
	\begin{tabular}{ccccccccccccc}
          % \toprule
    & n$_{\mathrm{DWD,BPS}}$ & Galaxy model & $f_{\mathrm{GW}}>10^{-4}$ & \multicolumn{3}{c}{Resolved with~\LISA}  & \multicolumn{3}{c}{Well localised} & \multicolumn{3}{c}{Measured mass} \\ 
  & & & &   2yr & 4yr & 8yr           & 2yr & 4yr & 8yr  & 2yr & 4yr & 8yr           \\
%    \bottomrule
		\hline
	He-He& 51500      &  9.7$\times 10^6$    & 1.9$\times 10^7$         &    1900 &  3500 & 5900             &     60  & 200  &    450  &  0 &   10 & 400\\
  & 7$\%$    &     2$\%$          &  31$\%$ & 26$\%$&  31$\%$       &          31$\%$           &  6$\%$     &    $12\%$ & 16$\%$ &0$\%$ &1$\%$ & 13$\%$\\
    He-CO& 72000      & 2.9$\times 10^7$     & 2.5$\times 10^7$         &       3600 &    5600 & 8600           &    500 &   1000 & 1700      & 100 &  600 & 1700 \\
	     & 10$\%$    &6$\%$               & 40$\%$ & 51$\%$ & 48$\%$ &  46$\%$          &     58$\%$  &  59$\%$ &  59$\%$ & 49$\%$  & 62$\%$ & 60$\%$   \\
	CO-CO& 609000      &4.4$\times 10^8$        & 1.2$\times 10^7$         &      1400 &   2200 & 3400            &   300 & 500 & 600         &  100 &   300 & 650   \\
	     &80 $\%$   &    87$\%$            & 19 $\%$&  19$\%$ & 18$\%$  &          18$\%$           &  31$\%$     &    26$\%$  & 22$\%$ & 42$\%$ & 32$\%$  & 24$\%$ \\
	Ne+X &30000       &    3.1$\times 10^7$  & 5.8$\times 10^6$         &        200 &  350 & 800           &  40 &       50 & 90      &  20 &  40 & 90\\
	     &4$\%$     &        6$\%$         & 9$\%$ & 3$\%$ & 3$\%$ &  4$\%$                    &      4$\%$               &   3$\%$      &    $4\%$ & 9$\%$ & 4$\%$  & 3$\%$\\
	Total & 763000    &    5.1$\times 10^8$  & 6.2$\times 10^7$         &      7000 & 12000          &    19000 & 900 & 1800 & 2800          & 200 &    1000 & 2800 \\
		\hline
		    & 1 & 2 & 3 & 4& 5 & 6& 7 & 8 & 9 & 10 & 11 &12 \\
		    \hline
	\end{tabular}
	
    \caption{Summary of the main properties of the different types of DWDs. Columns are numbered at the bottom. From left to right, we specify the approximate number of systems formed in the binary population model (initial population of 2.5 million systems at Solar metallicity, see \S\ref{subsec:BPS}), the corresponding present-day galactic population and the number of systems \fGW$>10^{-4}$ Hz.  Then we show the number of individually resolved systems, the number of well-localised systems and the number of systems with chirp mass measured within 10 per cent, for 2, 4 and 8 years of observations. We indicate the total number as well as the fraction of each subtype.}
	\label{tab:summary_numbers}
\end{table*}

In this paper, we will consider each population separately as they have different GW properties and we will show they also stem from different stellar populations.

Once the DWD binary is formed, we assume its evolution is only determined by GW emission. As such, systems born with short orbital periods merge before the present day and we remove them from our sample. Depending on the masses of the systems, such mergers may lead to type Ia supernovae~\citep{Iben_84_DWD_SNIa} or other transients~\citep{Saio_14_accretionDWD,Shen_12_DWD_mergers}. In some binaries, mass transfer may occur before the merger if at least one star fills its Roche lobe. For He-He DWDs, which have the largest radii, this typically occurs for periods below 3 minutes. The resulting mass transfer is unstable if the mass ratio of the binary $M_{\mathrm{donor}}/M_{\mathrm{acretor}}$ > 2/3, leading to  a fast merger. This means that all He-He and CO-CO DWDs will merge quickly, potentially on faster timescales than the GW emission predicts. The fate of binaries with smaller mass ratios (typically He-CO binaries) is less clear, the stability of the mass transfer depends on the spin-orbit coupling of the binary and the geometry of the accretion stream. Stable mass transfer could then lead to the widening of the orbit, which would keep the system in the~\LISA~band for longer \citep{Marsch_04_DWD_mass_transfer,Gokhale_07_DWD_contact}. Such systems would appear like AMCVn systems due to the presence of the accretion disk. We choose not to model these systems, and any other type of AMCVn system. According to ~\citet{Brown_2016_DWDmergers} most of these systems actually merge quickly, based on a study of the ages of a sample of Extremely Low Mass DWDs ~\citet[ELM][]{Brown_10_ELM_1,Brown_16_ELM_VizieRcat}, and only a small fraction become an AMCVn binary.  This was also predicted by \citet{2015ApJ...805L...6S} who proposed that even accreting double WD binaries with extreme mass ratios will merge due to classical nova-like outbursts on the accretor. As such, we only account for orbital evolution due to GW emission in this paper. 

\subsection{Combined binary model and galaxy model}\label{subsec:combination}

Each star particle within $\simeq$ 300 kpc (roughly the virial radius of the Milky Way at $z=0$) gets assigned $n_{\mathrm{DWD}}$ white dwarf binaries, depending its stellar mass at birth $M_*$ and its metallicty (although the impact of metallicity is limited). We have
\begin{equation}\label{eq:num_DWD}
n_{\mathrm{DWD}}=\frac{M_*}{M_{\mathrm{tot,BPS}}}N_{\mathrm{DWD,BPS}},
\end{equation}
where $N_{\mathrm{DWD,BPS}}$ is the number of DWDs formed in a given binary population synthesis model resulting from an initial stellar population of total stellar mass $M_{\mathrm{tot,BPS}}$ (see column 1 in Tab.~\ref{tab:summary_numbers}). We only model $N_b$ binary systems with primary and secondary masses allowing them to form DWDs, representing a stellar mass of $M_b$. Although we reject $\bar{N}_b$ other binaries for the DWD modelling, their mass $\bar{M}_b$ should count towards ${M_{\mathrm{tot,BPS}}}$.  Assuming a binary fraction $f_b=0.5$, $M_{\mathrm{tot,BPS}}$ should also account for a total number of $N_b+\bar{N}_b$ single systems, which are drawn for a complete Kroupa IMF (between 0.1 and 100 $\msun$). In total, the equivalent stellar mass we model is given by
\begin{equation}\label{eq:total_equiv_mass}
M_{\mathrm{tot,BPS}}=\frac{1-f_b}{f_b}\sum\limits^{N_b+\bar{N}_b} m_1 +\frac{f_b}{f_b}\sum\limits^{N_b+\bar{N}_b}m_1+m_2.\\
\end{equation}
As such, our subsample of 2.5 million binaries represent a total stellar mass of roughly 5.1 10$^{7}$ $\msun$.

All the DWDs are stored in a dataframe. We randomly draw with replacement $n_{\mathrm{DWD}}$ from our BPS model for each star particle and add them to the dataframe.  The DWDs inherit the formation time and metallicity of the progenitor star as well as its current position and position at formation. The formation time of the DWD is the sum of the formation time of the progenitor and of the DWD. DWDs with formation times beyond the present day are removed (about 50$\%$ of the initial sample). We forward model the binaries until the present day via gravitational wave radiation, gradually shortening the orbit. We remove binaries that have already merged (less than 10$\%$ of the sample). If the binary has not merged yet, its semi-major axis $a$ and eccentricity $e$, which determine its GW emission, evolve according to \citep{Peters63}
\begin{eqnarray}\label{eq:orb_evol}
\frac{de}{dt}&=&-\frac{304}{15}\frac{G^3\mu (M_1+M_2)^2}{c^5a^4 }\frac{1}{(1-e^2)^{5/2}}\left(1+\frac{121}{304}e^2 \right)\\ \nonumber
\frac{da}{dt}&=&-\frac{64}{5}\frac{G^3\mu (M_1+M_2)^2}{c^5 a^3}\frac{1}{(1-e^2)^{7/2}}\left(1+\frac{73}{24}e^2+\frac{37}{96}e^4\right),\\ \nonumber
\end{eqnarray}
where $\mu$ is the reduced mass of the system. Typically one star particle generates 70-80 DWDs. To avoid spurious spatial clustering, we distribute the DWDs associated with each star particle relative to the particle's position using a spherical Epanenchnikov (quadratic) kernel whose length is determined adaptively based on the Mahlanobis distance to the $\sim 10$ nearest neighboring star particles, using the EnLink algorithm described in \citet{2009ApJ...703.1061S}. 

 Given our initial sample of DWDs, each DWD typically gets chosen for random assignment to a star particle about 200 times over the roughly 10 million star particles, with different positions and ages, representing the simulated galaxy. This combination guarantees that the final catalog of DWDs is only composed of truly unique binaries.

\subsection{Modeling the gravitational wave emission}\label{subsec:GW_methods}
To estimate the capability of~\LISA~to detect and characterize white dwarf binaries in the galaxy models, we simulate the~\LISA~data by co-adding the gravitational waveforms from all binaries with signals in the measurement band using the fast waveform generator in \cite{Cornish:2007}.
Without replicating the derivation, it is valuable to point out here that the dimensionless GW strain from a compact binary at a distance $r$ is given by
\begin{equation}\label{eq:GW_strain}
    h=2(4\pi)^2f_{\mathrm{GW}}^{2/3}\frac{G^{5/3}}{c^4}\frac{M_c^{5/3}}{r}.
\end{equation}
The measurement of the GW strain and frequency alone are insufficient to determine the chirp mass of the binary, which is degenerate with the distance. The chirp mass (and therefore distance) can be determined for galactic binaries in the~\LISA~band, having wide orbital separations and orbital velocities $\ll c$, to leading order in the frequency evolution 
\begin{equation}\label{eq:GW_Mchirp_fdot}
    M_c=\frac{c^3}{G}\left(\frac{5}{96}\pi^{-8/3}f_{\mathrm{GW}}^{-11/3}\dot{f}_{\mathrm{GW}}\right)^{3/5}
\end{equation}
assuming that other contributions to the orbital period evolution (e.g. mass transfer, tides, etc.) are sub-dominant effects. Note that $\dot{f}_{\rm GW}$ is a difficult parameter for~\LISA~ to constrain, so chirp mass measurements are only possible for ``outliers'' of the total population, requiring high signal-to-noise (\SNR), comparatively large $\dot{f}$, and/or long integration times for the~\LISA~observations.
Fortunately, due to the large number of detectable binaries, even the tails of the source distribution are well populated.

Our simulated~\LISA~response to the galaxy models use spacecraft noise levels and orbits consistent with those in the~\LISA~mission proposal~\citep{LISA_proposal}. 
Determining which of the simulated binaries are individually resolvable is challenging in the regime where the confusion noise dominates the data stream.
From the full input binary population, the detectable sources are identified by using an iterative process that utilizes a median smoothing of the power spectrum to estimate the noise level (dominated by source confusion around 1 mHz), regresses binaries from the data with signal-to-noise ratios \SNR$>7$ as detected sources, and cycles until the detection ``catalog'' converges~\citep{Timpano:2006,Nissanke_2012_DWD}. 
This approximate method is in qualitative agreement with more realistic search strategies~\citep{Crowder:2007, Littenberg:2011}

For the detected binaries, we assume Gaussian measurement uncertainties centered on the true parameters for each source, completely characterized by the covariance matrix. To compute the inverse covariance matrix we use the Fisher approximation, with a central differencing numerical differentiation scheme to calculate derivatives of the waveforms~\citep{Cutler:1994}. 
The Fisher matrix is only an approximation with well-publicized shortcomings~\citep{Vallisneri:2008} but, similar to the argument for the hierarchical search method, a more robust error analysis (e.g., using stochastic sampling methods) requires prohibitive computationally resources for the scope of this work.
Furthermore, where the Fisher approximation is most accurate is in the high \SNR\ regime, and many of the results in this work are focused on exactly those binaries because they are the systems that yield the best parameter constraints.

The computation of the GW signal is only performed for the binaries with present-day frequency above $10^{-5}$ Hz.

\section{Results}\label{sec:results}

\subsection{Origin of different populations of short-period DWDs in the Milky Way}\label{sec:result_MW_model}
Here we describe the properties of the tight DWD binaries in a Milky-Way like galaxy based on the combination of the cosmological simulation and the binary population synthesis model. We will describe how different DWD populations arise from the global stellar population and vice versa.

\begin{figure*}
	\includegraphics[width=.4\textwidth]{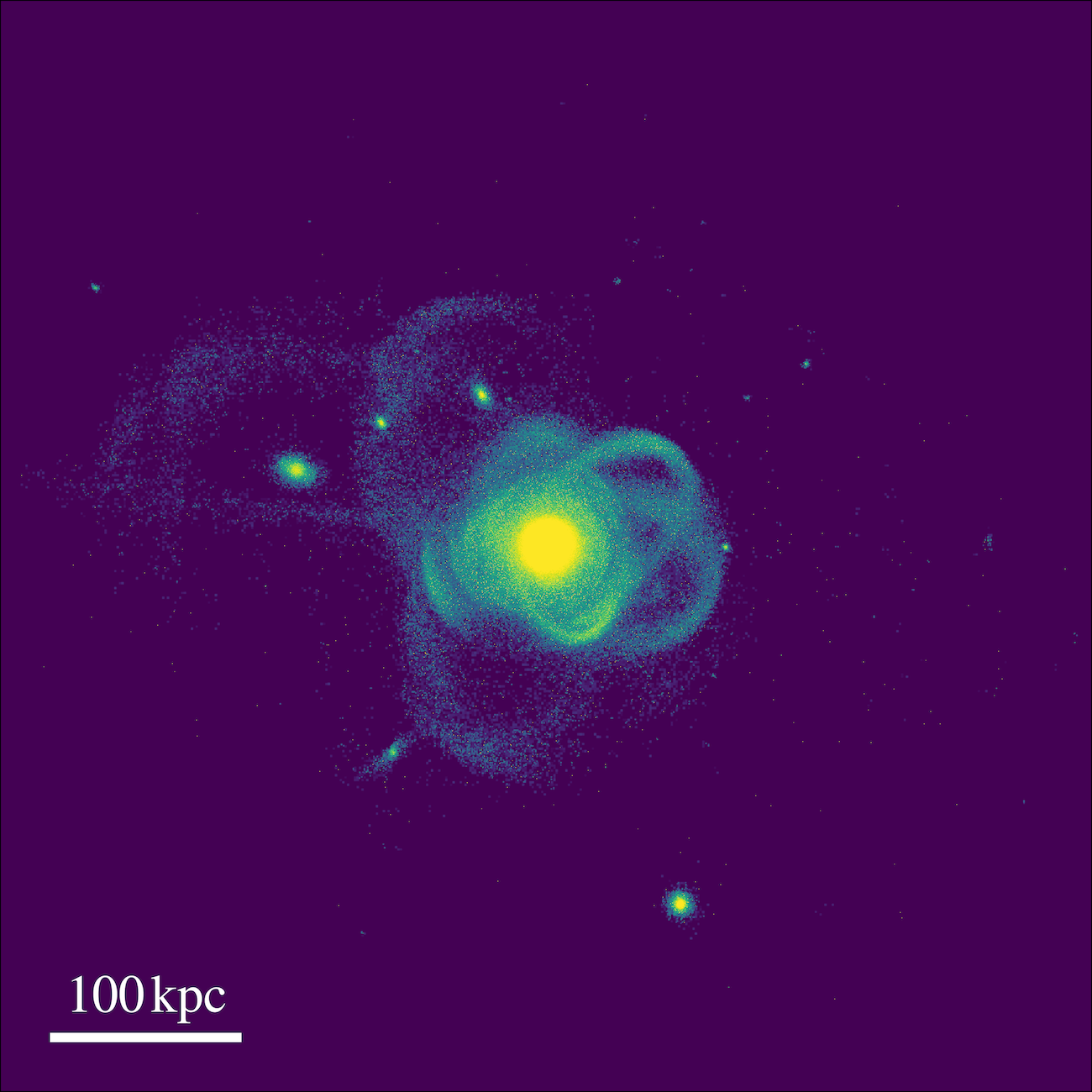}
	\includegraphics[width=.4\textwidth]{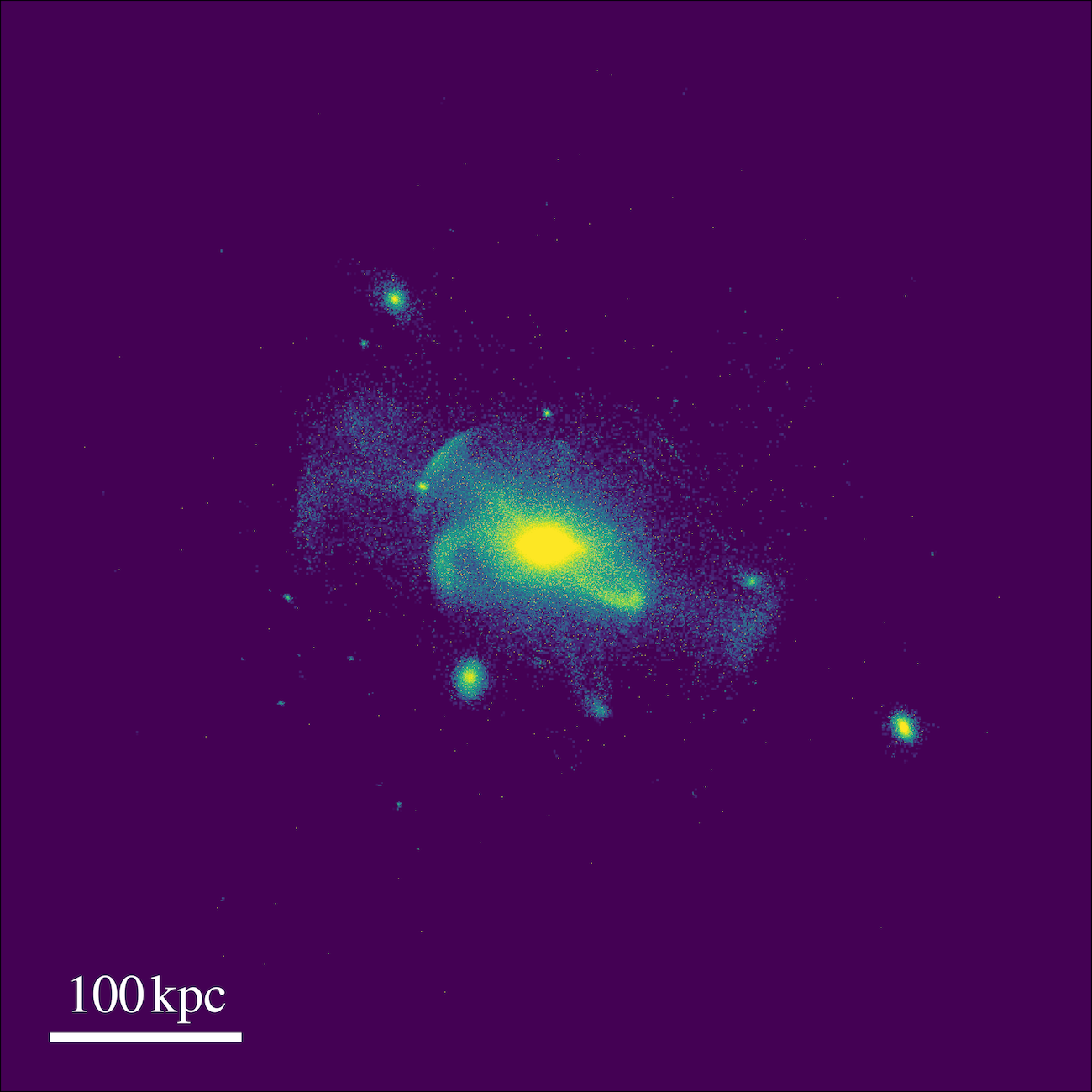}
	\includegraphics[width=.095\textwidth]{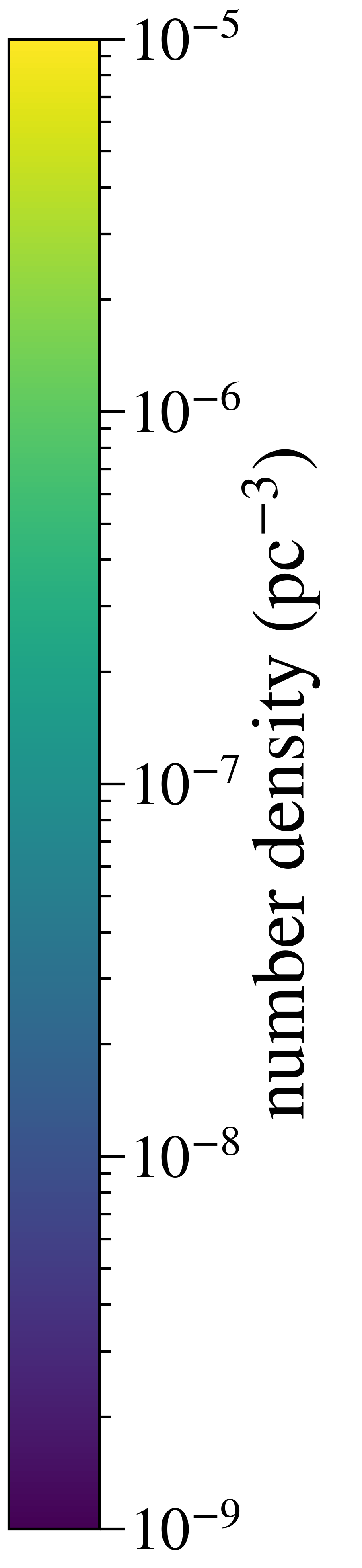}
    \caption{Large-scale maps of the surface density of the number of DWDs in the simulation, viewed faced on (left) and edge-on (right). To zeroth order, the spatial distribution is very similar to the stellar distribution (not shown). The main galaxy shows a disk, bulge and stellar halo. DWDs are also present in the tidal streams and satellite galaxies.}
    \label{fig:maps}
\end{figure*}

The second column in Tab.~\ref{tab:summary_numbers} shows the total numbers of the DWDs in and around our MW-like Galaxy. About 500 million systems have formed, most of them at very wide orbits. In comparison with the direct output from the binary evolution model (which assumes a unique burst of star formation) we find that He-He and He-CO DWDs are under-represented in the Galaxy model. This is because many of these DWDs form at high frequencies, and many have merged by the present-day. Additionally, for the He-He DWDs,  the formation time is comparatively longer, which means the recently formed stars can not have contributed to this population. Of the 500 million binaries in the MW, most of them are too separated to have ever interacted. In the remainder of  this paper, we will only focus on binaries that are relevant for~\LISA.

The third column in Tab.~\ref{tab:summary_numbers} shows the number of systems with a $f_{\mathrm{GW}}>10^{-4}$ Hz. There are roughly 60 million systems in our simulation, which is consistent with previous results~\citep[e.g.][]{nelemans01_WD,Ruiter_2010_DWD,Nissanke_2012_DWD}. He-He and He-CO DWDs are over-represented at these high frequencies (respectively 31 and 40 per cent) because they have undergone two common-envelope phases. Fig.~\ref{fig:maps} shows the surface density of the short-period binaries in and around the galaxy, with and edge-on and face-on view. Overall, the distribution is similar to the stellar distribution (not shown here). As our galaxy model is based on a cosmological simulation, the DWDs are naturally present in all the components of the Milky Way such as the thin and thick disk, the bulge,  the stellar halo, tidal streams and satellite galaxies. This is different from all previous models where the spatial distribution is a parametrized, symmetrical model of the disks and bulge (and sometimes the stellar halo, as in ~\citet{Ruiter_2010_DWD}).

\begin{figure*}
	\includegraphics[width=.45\textwidth]{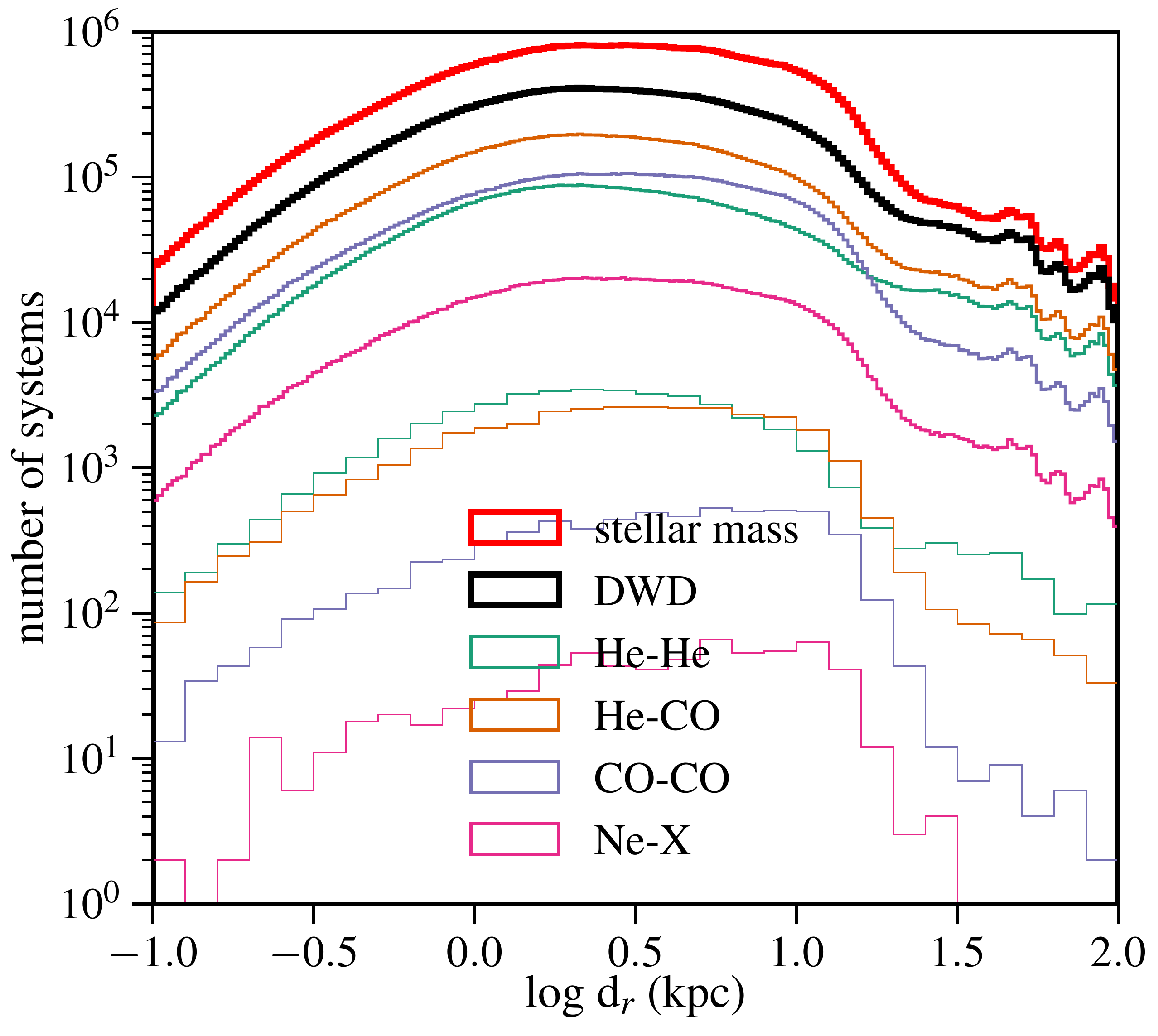}
	\includegraphics[width=.45\textwidth]{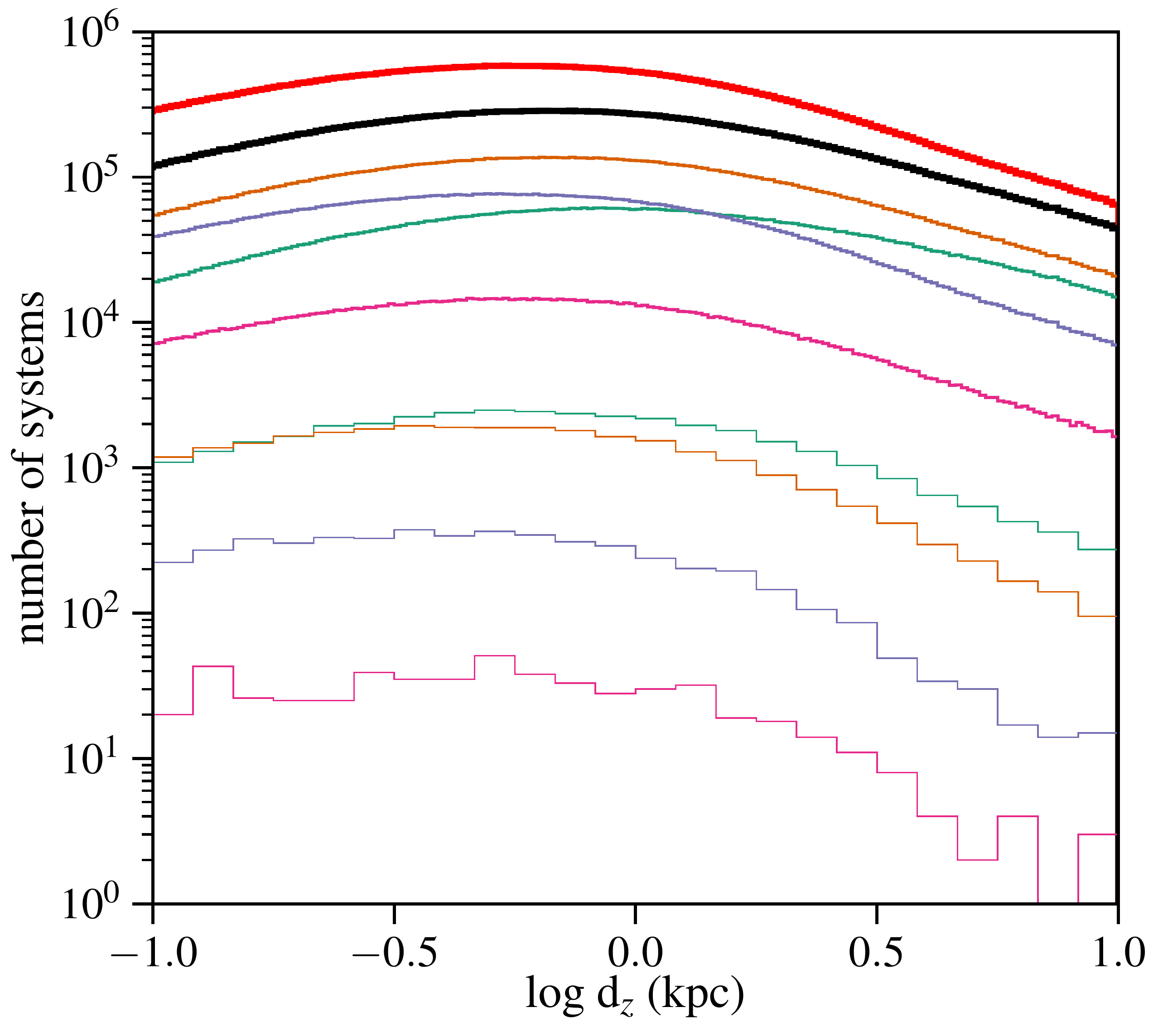}
    \caption{Spatial distributions of the DWDs (black) in comparison with the stellar mass (red, rescaled by a factor 1000). Different colors show different subtypes, and the thin lines only show the binaries with $f_{\mathrm{GW}}>10^{-3}$ Hz. Both plots have the same legend. }
    \label{fig:hist_dist}
\end{figure*}

Fig.~\ref{fig:hist_dist} shows a histogram of the radial distance with respect to the galactic center (left) and the distance above the plane (right) of the binaries. Globally, the DWD distribution follows the stellar distribution, although DWDs somewhat prefer the bulge, stellar halo and satellites and are less present in the disk. This effect is mostly visible for CO-He DWDs and even more He-He DWDs, which become more numerous than the CO-CO DWDs outside of the disk. On the contrary, the DWDs stemming from the most recent stars, such as CO-CO and Ne-X DWDs slightly prefer the disk. This trend is confirmed in the other cosmological simulations we analysed. These distributions can be explained by the minimal stellar age of each population (see Fig.~\ref{fig:prog_props}). 

\begin{figure*}

	\includegraphics[width=.305\textwidth]{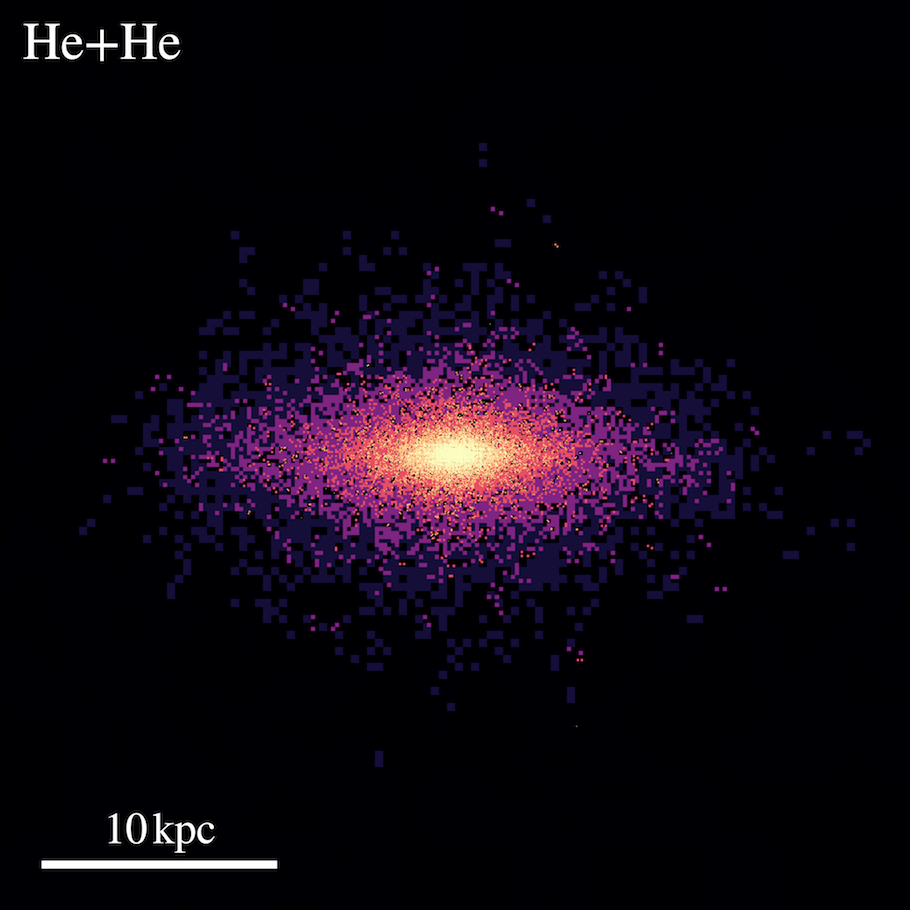}
	\includegraphics[width=.305\textwidth]{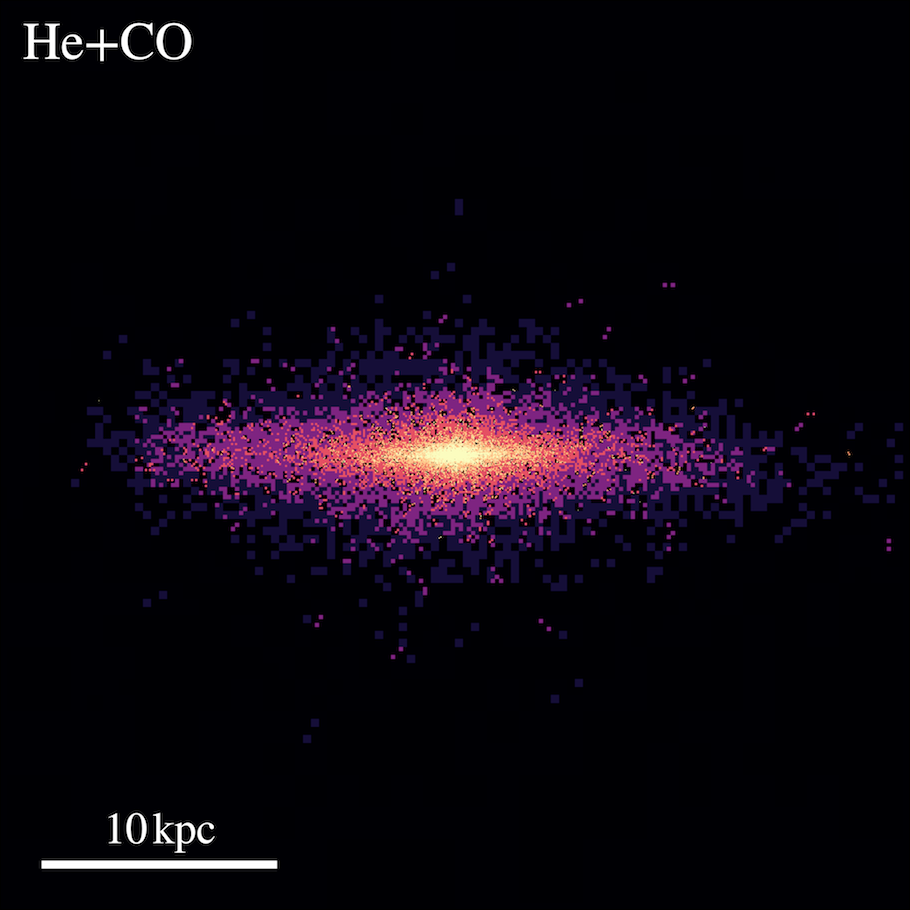}
	\includegraphics[width=.305\textwidth]{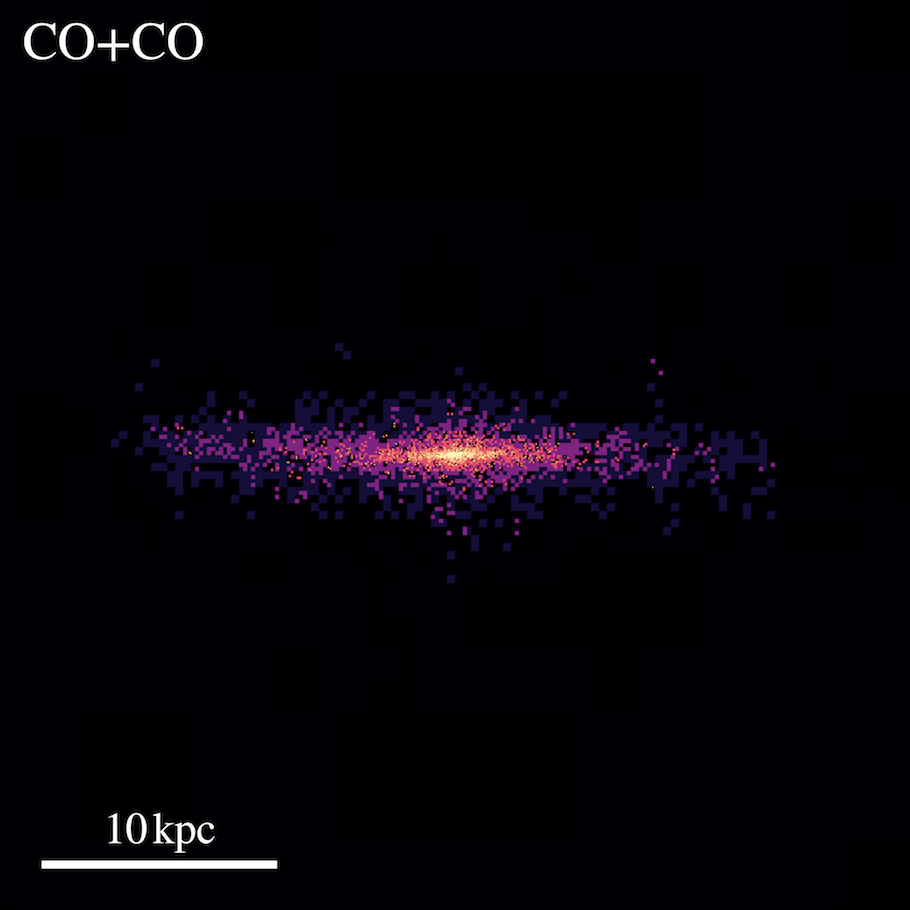}
	\includegraphics[width=.058\textwidth]{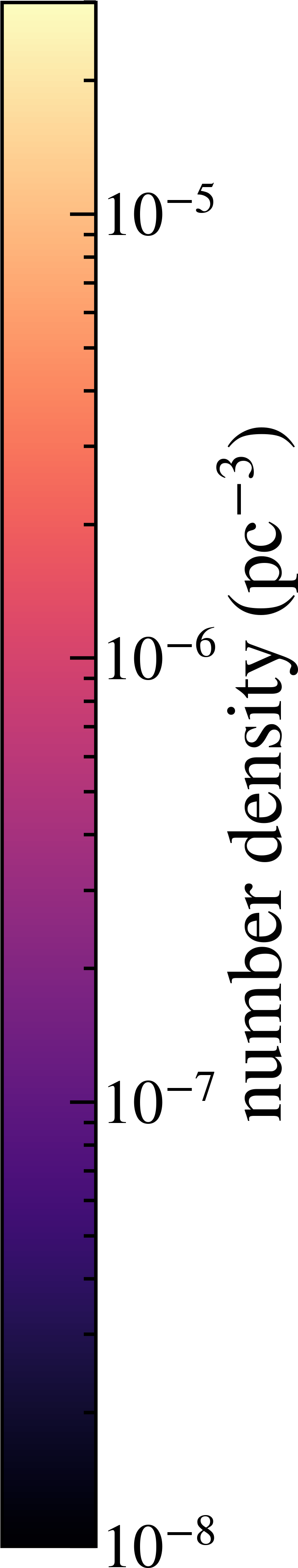}
    \caption{Maps of the He-He DWDs (left), He-CO (middle) and CO-CO DWDs (right) viewed and edge-on. These maps only show the binaries with \fGW$>10^{-3}$ Hz, which are the most likely to be individually resolved by~\LISA. We do not show the Ne+X binaries due to their sparcity. Their distribution is very similar to the COCO binaries.}
    \label{fig:maps_hf}
\end{figure*}

When focusing only on the highest frequency systems shown with thin lines ($f_{\mathrm{GW}}>10^{-3}$ Hz, where most of the individually resolved sources for~\LISA will be found), we find that they more closely follow the stellar distribution than their low-frequency counterparts. Still, they keep distinct spatial distributions, as is shown in Fig.~\ref{fig:maps_hf}. These maps show the galaxy edge-on: the He-He DWDs distribution is almost spherical due to the bulge and halo, with a thick disk. On the opposite, the CO-CO DWDs are present almost exclusively in a very thin and elongated disk.  The He-CO DWDs present an intermediate distribution, with prominent disk, although with a smaller scale height than for He-He DWDs, and a limited contribution from the bulge and halo.

\begin{figure}
	\includegraphics[width=.45\textwidth]{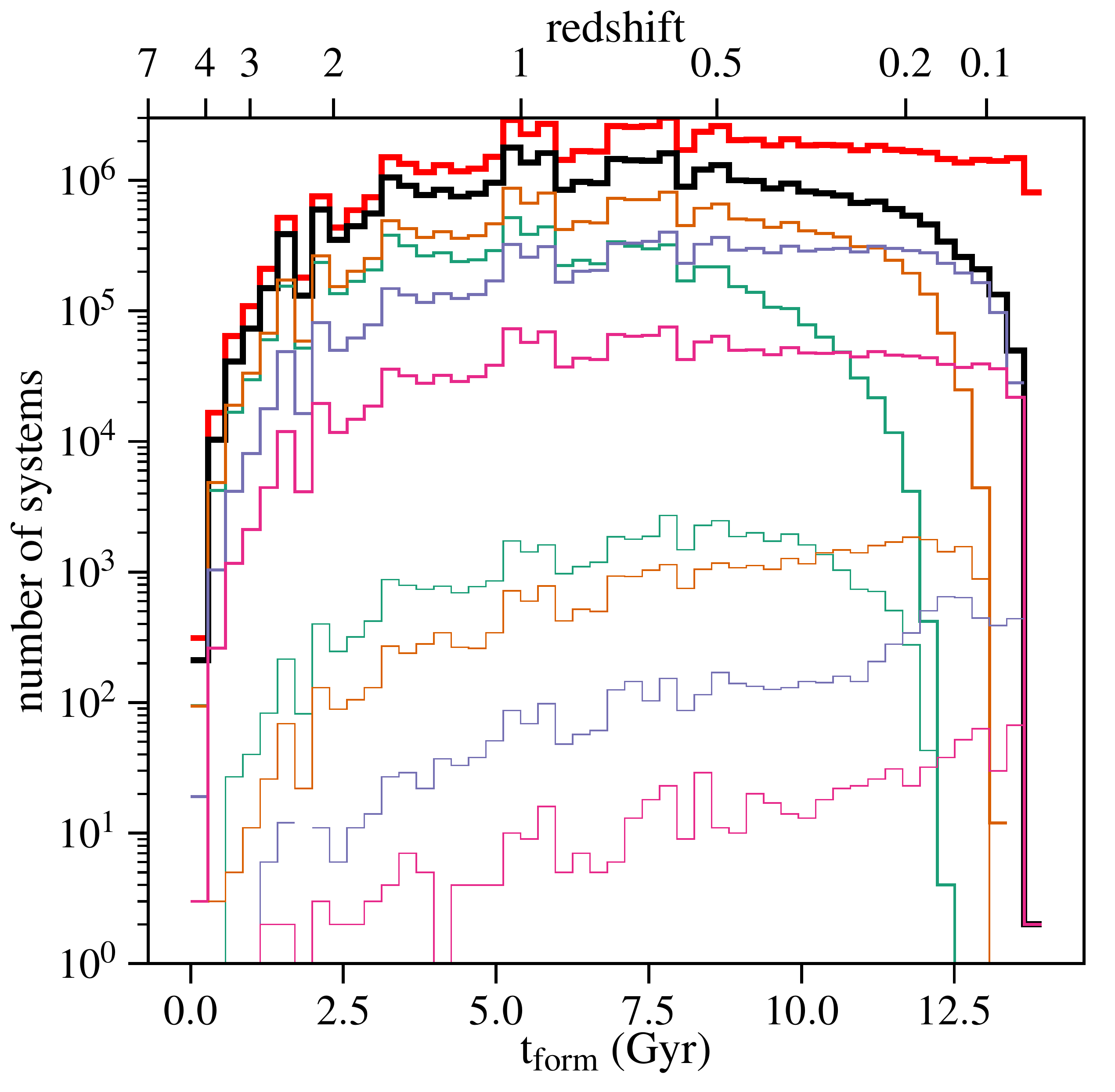}
    \caption{Formation time of the progenitors of the DWDs (black) in comparison with the stars (red, rescaled). Different colors show different subtypes, and the thin lines only show the binaries with \fGW $>10^{-3}$ Hz (colorscheme is identical to Fig.~\ref{fig:hist_dist}).}
    \label{fig:hist_tform}
\end{figure}

A more complete understanding of the present-day DWD population in a MW-like galaxy comes from the formation time of their progenitor stars. Fig.~\ref{fig:hist_tform} shows that the formation of DWD progenitors follows the global star formation rate until $z\simeq 1 $ where it starts declining. This is related to the typical formation time of 2 Gyr for a CO-He DWD and beyond 5 Gyr for a He-He DWD. Conversely CO-CO and Ne-X DWDs can form on a much shorter timescale and trace their progenitors star formation history almost completely. For the high frequency systems, only young DWDs are present, so the contribution increases towards recently formed progenitors, except for the He-He DWDs. Again, the different behaviour for He-He DWDs is the wider range of the duration of stellar evolution. A He-CO, CO-CO or Ne-X DWD detected with high frequency probably stems from a progenitor formed less than 2 Gyrs ago (z$\lesssim 0.2$), while a He-He DWD progenitors likely formed between 3.5 and 6 Gyrs ago. These effects explain the different spatial distributions shown in Fig.~\ref{fig:maps_hf}.
As the metallicity of the galaxy globally increases with time, we find that CO-CO and Ne-X DWDs stem from stars with the same metallicity distribution as the global stellar population, peaking around $Z=3\Zsun$ while the He-He DWDs mostly stem from stars with lower metallicity (peaking around $\Zsun$).

\subsection{Prospects for detecting short-period DWDs in the Milky Way}\label{sec:results_detectability}
\begin{figure}
	\includegraphics[width=.45\textwidth]{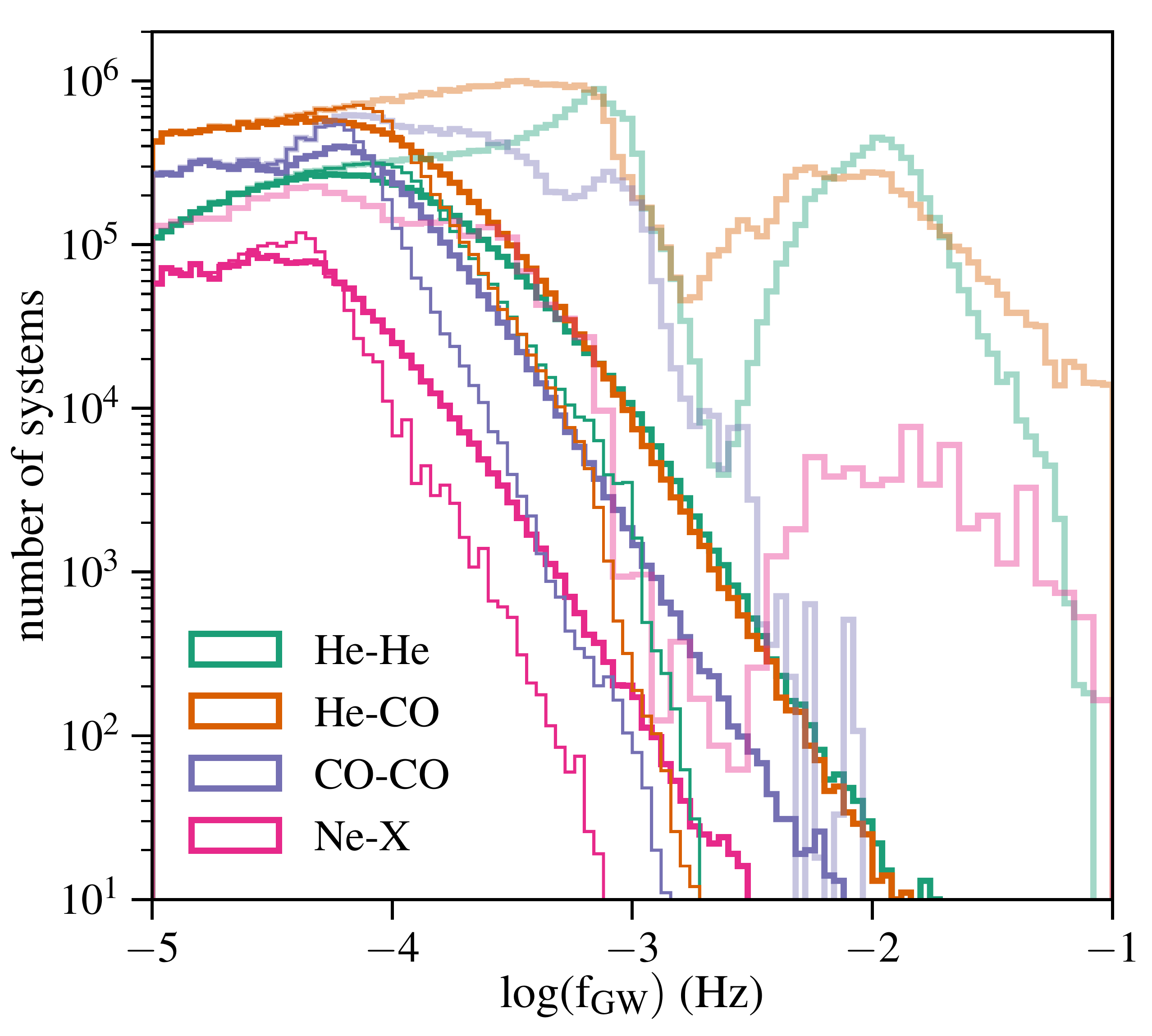}
    \caption{Distribution of the frequency of the DWD binary population. We only show binaries with \fGW$>10^{-5}$ Hz, which are in the~\LISA~frequency band. The fainter lines show the frequency distribution at the formation of the DWD at formation and the thin lines show the initial frequency of the binaries which have not merged by the present day.}
    \label{fig:hist_freq}
\end{figure}

Here we discuss the frequency/orbital period distribution and distance to the Sun of the different binary populations. These are crucial aspects for detection with gravitational waves and/or electromagnetic observations. Fig.~\ref{fig:hist_freq} shows the frequency distribution of all the binaries with \fGW$>10^{-5}$ Hz. The CO-He DWD population dominates below $\simeq 10^{-3}$ Hz and the He-He DWDs are slightly more numerous beyond that. Initially roughly 90 million binaries were formed with \fGW$>10^{-5}$ Hz (faint lines), with their maximal frequency up to a few Hz. All the binaries with initial \fGW$>10^{-2}$ Hz have merged by now. For binaries formed with \fGW$>10^{-5}$ Hz, 40$\%$ are still currently present, with 32\%, 44\%, 47\%, 20\% of the He-He, He-CO, CO-CO, Ne-X binaries still present, respectively. He-He and Ne-X DWDs undergo most of the mergers because of their tight initial orbits (for He-He DWDs) and their high chirp masses (for Ne-X DWDs). Note that we have not removed He-He binaries undergoing Roche Lobe overflow from this sample (see \S\ref{subsec:BPS}). The overall low number of  high-frequency CO-CO DWDs is due to their limited birth rate at high frequency (see shaded area in Fig.~\ref{fig:final_Mtot_v_P}) combined with a  merger time of a few tens of millions of years for systems initially formed with an orbital period of one hour. In comparison He-He DWDs with an initial period of an hour need about 100 million years to merge. In all cases, the DWDs in high frequency systems must have formed recently, otherwise they would have merged. Given the wide range of formation times of He-He DWDs, recent DWD formation does not impy that the progenitor binary formed recently.

The initial frequency distribution of the currently present DWD binaries (thin lines) shows that most of them initially had a frequency around $10^{-4}$ Hz and have hardened to their present-day orbital frequencies. Systems with initial frequency below $\simeq 5 \times10^{-5}$ Hz have not evolved significantly since their formation. This implies that most DWD binaries, which are formed with a lower frequency, will never be relevant to~\LISA.

\begin{figure}
	\includegraphics[width=.45\textwidth]{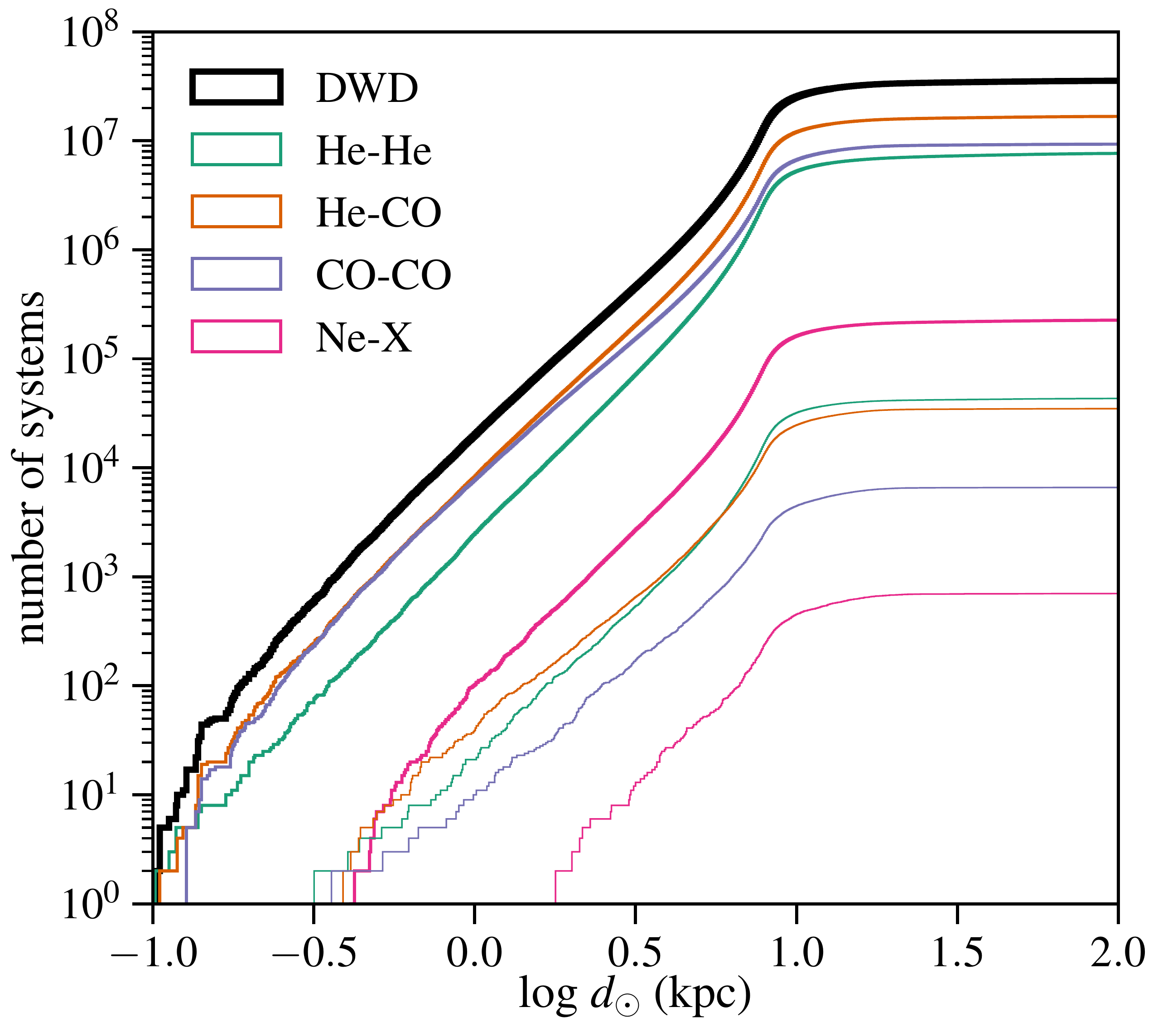}
    \caption{Cumulative distribution of the distance to the Sun of DWDs (black). Different colors show different subtypes, and the thin lines only show the binaries with \fGW$>10^{-3}$ Hz.}
    \label{fig:CDF_dist_Sun}
\end{figure}

Fig.~\ref{fig:CDF_dist_Sun} shows the cumulative distribution of the DWDs within a certain distance to the Sun. CO-CO and He-CO binaries dominate at all distances. When focusing only on the highest frequency systems, CO-He DWDs are the most numerous within a few kpc, but there is a significant contribution of He-He and CO-CO DWDs as well. This is roughly the distance up to which \textit{Gaia} will be able to observe verification binaries. Beyond $\simeq$ 5 kpc, which will be observable by LSST, He-He and He-CO DWDs largely dominate the sample. In the following section we present the observable properties of the GW emission of our population.

\section{GW signatures}\label{sec:GW}
In this section we predict the detections of the short-period DWDs with~\LISA, based on the complete GW emission model described in \S\ref{subsec:GW_methods}. We explain how the binaries described in \S\ref{sec:results_detectability} are affected by~\LISA's response function.  We present the properties of the individually resolved sources (\S\ref{subsec:LISA_resolved}), including implications of mass and distance measurements as well as sky localisation. As our method includes a cosmological model of the galaxy, we detail the possibility to detect and identify sources in the stellar halo, stellar streams and satellites (\S\ref{subsec:GW_halo}).

\subsection{Individually resolved sources}\label{subsec:LISA_resolved}
Columns 4-6 in Table \ref{tab:summary_numbers} summarise the number of sources individually resolved by~\LISA~over time. A source is considered to be resolved if it can be uniquely identified within its frequency bin, with a \SNR of at least 7. Contrary to (most) electromagnetic detections, this definition of a source does not have any implication on our ability to localize it on the sky. We find roughly 12000 resolved binaries after a 4-year mission. 

\begin{figure}
	\includegraphics[width=.45\textwidth]{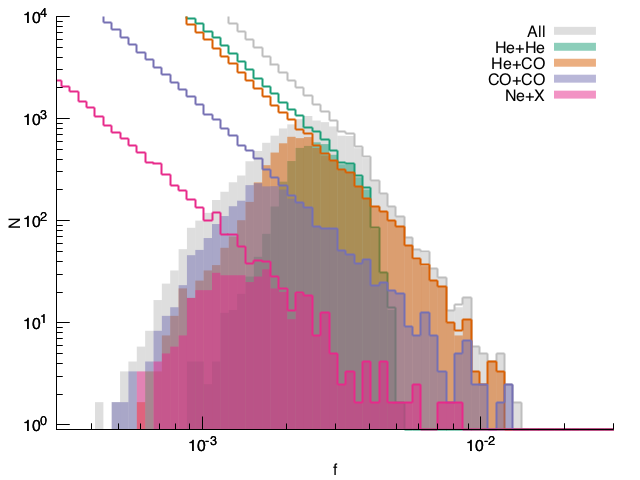}
    \caption{Frequency distribution of the systems with \SNR>7 in comparison with the total distribution after an 8 year observing time. Note that the catalog of detected binaries with $f_{\rm GW}\gtrsim3$ mHz ($f_{\rm GW}\gtrsim2$ mHz for the more massive UCBs), is complete.}
    \label{fig:hist_freq_LISA}
\end{figure}

Fig~\ref{fig:hist_freq_LISA} shows the histogram of the frequency of the different types of binaries detected by~\LISA. This plot directly shows how the global frequency distribution in Fig.~\ref{fig:hist_freq} translates into~\LISA~detections. We find that He-CO DWDs and and He-He DWDs are the most numerous in the~\LISA~band and among the detected sources, even though their contribution to the global galactic population is about 5$\%$ at most. This is because these binaries typically have the tightest orbits. He-He DWDs are present only up to about 5 mHz, because He WD have the largest radii. At higher frequencies, Roche Lobe overflow will happen and mass transfer will quickly become unstable and lead to a merger. As such, we remove binaries above this frequency from our sample (see end of \S\ref{subsec:BPS}). 

\begin{figure*}
	\includegraphics[width=.24\textwidth]{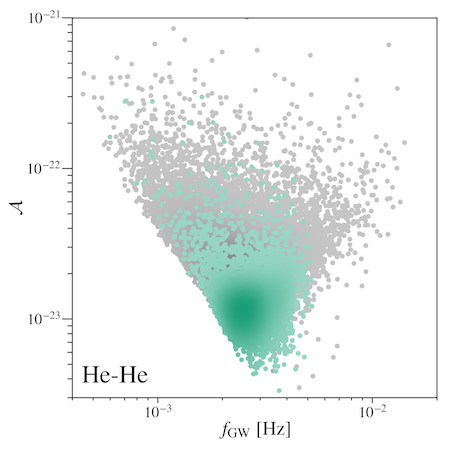}
	\includegraphics[width=.24\textwidth]{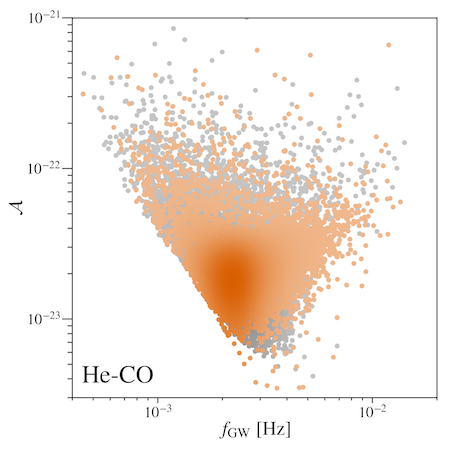}
 \includegraphics[width=.24\textwidth]{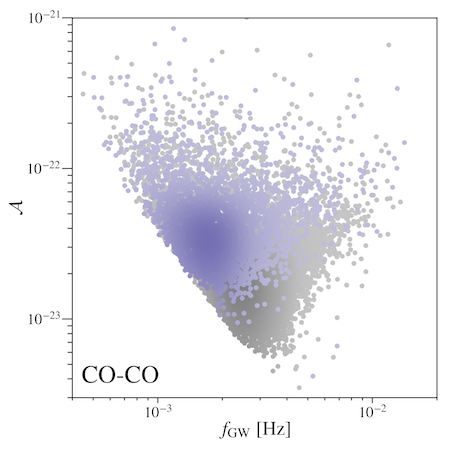}%
	\includegraphics[width=.24\textwidth]{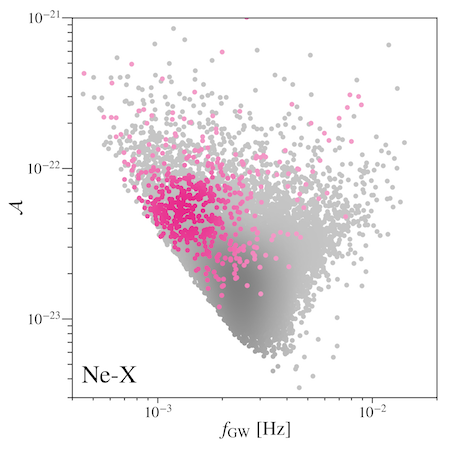}
    \caption{Heat maps of gravitational wave amplitude versus frequency for all the systems with \SNR$\geqslant$ 7. The color schemes are linear and there is less than an order of magnitude difference between the brightest spots and the faintest. }
    \label{fig:hist_GW_amp_LISA}
\end{figure*}

Fig.~\ref{fig:hist_GW_amp_LISA} shows the contribution of the individually resolved systems of each  type of binary to the GW signal.  These maps show the same frequency dependence as Fig.~\ref{fig:hist_freq_LISA}. The GW amplitude is set by the distance to the source, the frequency of the binary and the chirp mass (Eq.~\ref{eq:GW_strain}). The He-He systems are very numerous but have the lowest GW amplitude due to the lower chirp mass, followed by the He-CO systems. Conversely the CO-CO and Ne-X systems are less numerous but contribute at higher amplitudes.

Naturally the number of detected systems will increase over time as signal is accumulated and more and more systems become detectable above the instrumental and confusion noise. The different columns in Tab.~\ref{tab:summary_numbers} emphasize that particularly the He-He DWDs will mostly benefit from an extended mission. After the nominal mission of 4 yrs, a third of the newly detected systems will be He-He DWDs, as opposed to one quarter during the first 2 years of the mission. This is due to the fact the He-He DWDs have a low GW amplitude (see Fig.~\ref{fig:hist_GW_amp_LISA}) and a low frequency (see Fig.~\ref{fig:hist_freq}) which means that many of them will be buried in the foreground noise, which will decrease as the number of resolved systems increases. The determination of the frequency of the systems will allow statistical studies of binary evolution.  The comparison between the resolved population (solid histogram in Fig.~\ref{fig:hist_freq_LISA}) and the complete DWD population in the galaxy (coloured lines) highlights that the sample of resolved binaries is complete down to 3 mHz, and even 2mHz for the most massive binaries. Effectively, any binary with a period below 15 minutes will be individually resolvable, no matter its location in our galaxy, including the nearby satellites. As such, all the detections of these systems will be crucial to constrain binary evolution.

\begin{figure}
    \centering
    \includegraphics[width=.45\textwidth]{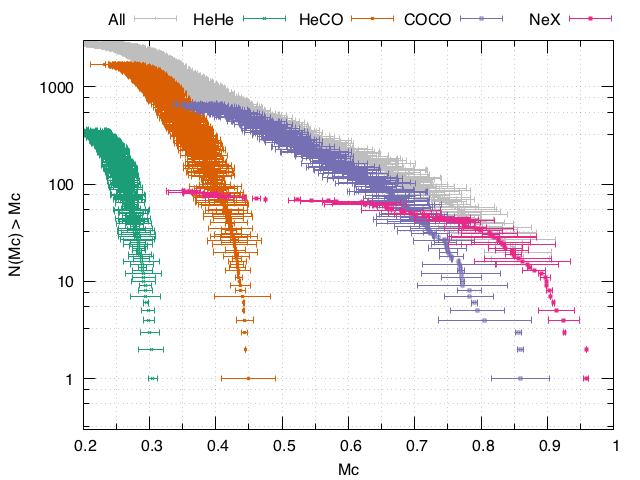}
    \caption{Survival functions of binaries with chirpmass determined to better than 10\%. For this simulation (08 year observing time) we can expect to unambiguously identify O(1000) He-CO DWDs, O(300) CO-CO DWDs, and O(10) Ne-X systems.}
    \label{fig:cdf_Mc_08}
\end{figure}

For most of the sources, the measured frequency of the binary will be effectively constant during the observing time. For certain sources,~\LISA~will also be able to measure the first frequency derivative $\dot{f_{\mathrm{GW}}}$ and determine both the chirp mass and distance of the binary (Eqs.~\ref{eq:GW_strain}-\ref{eq:GW_Mchirp_fdot}). This will only occur for chirping binaries, which frequency changes over the course of~\LISA's lifetime:  the systems with the highest masses and frequencies.  Fig.~\ref{fig:cdf_Mc_08} shows the binaries with chirp masses measured with better than 10 per cent uncertainty after an extended mission of 8 years. Out of the $\simeq$ 3000 systems the majority will be CO-CO ($\simeq$ 700) and He-CO DWDs ($\simeq$ 1700). The detailed numbers are provided in columns 10-12 of Tab.~\ref{tab:summary_numbers}. 

Given the complexity of stellar evolution, DWDs of a given subtype can have a wide range in chirp masses. This limits the classification of DWDs based on the chirp mass to a few hundred systems. This would probably also be quite dependent on the details of the binary model. Unfortunately~\LISA~will not be able to determine the component masses of DWDs, making the classification of the different binaries very uncertain and additional information from EM observations or theoretical models may be necessary.  

 Mass measurements become possible when a frequency derivative is measured. The latter really benefits from long integration times for the observations. Globally, during the first 2 years, 200 systems will have mass measurements, the 2 following years will yield an additional $\simeq$ 750 measurements and an extended mission would yield about 450 additional measurements per year. These values are crucial to characterise the systems and most of the statistics will  be obtained with an extended mission. This is even more the case for He-He DWDs, which will only contribute to the mass measurements during an extended mission. The latter are ideal candidates for EM observations due to their large radii.

\begin{figure}
\includegraphics[width=.45\textwidth]{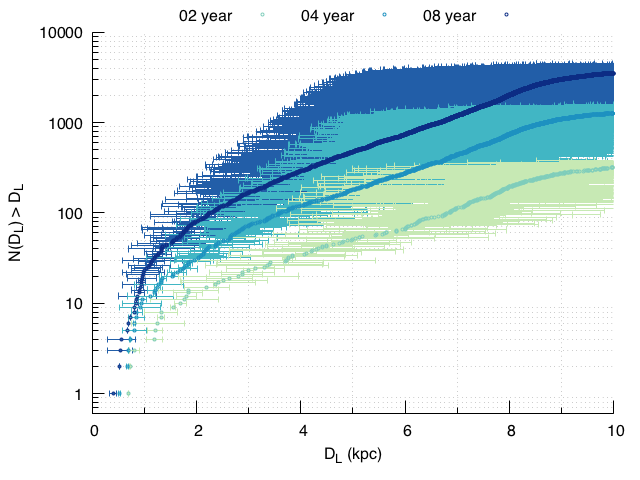} 
    \caption{Distance measurements and their uncertainties of all the well-localised binaries within 10 kpc from the Sun after 2, 4 and 8 yrs of the~\LISA~mission.}
    \label{fig:dist_resolved_10kpc}
\end{figure}

Sky localisation and distance measurements are crucial to enable the identification of electromagnetic counterparts, based on existing data or new observations. A source is considered to be well localised if we can determine its position in the sky within 10 square degrees and its distance within 50 per cent. These sources are the best candidates for EM follow-up or cross-matching with EM catalogs. Columns 7-9 in Tab.~\ref{tab:summary_numbers} indicate the number of well-localised systems in the simulation.  Current large scale sky surveys reach an average depth of $\approx$20-21\,mag which limits surveys to $\approx$1-2\,kpc. In a few years from now LSST will reach an average $r$-band depth of $\approx$24.5\,mag in a single epoch and $\approx$27.5\,mag in the co-added map \citep{2019ApJ...873..111I} allowing us to detect the electromagnetic counterpart up to $\approx$5 and $\approx$10\,kpc respectively. Fig.~\ref{fig:dist_resolved_10kpc} shows the expected number of systems with distance uncertainties of 50\,\% and sky localisation better than 10\,square degrees, well matched to LSST's field-of-view. At the end of the nominal mission, a few hundred systems could have counterparts. With an extended mission and stacked LSST data, a few thousand systems could be found.

\begin{figure*}
\includegraphics[width=.45\textwidth]{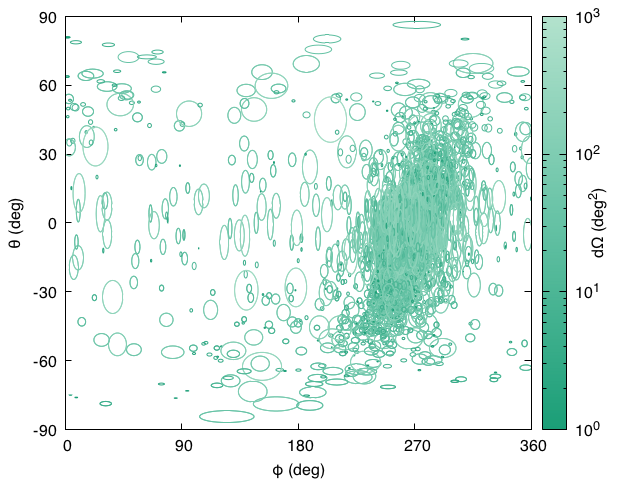} 
\includegraphics[width=.45\textwidth]{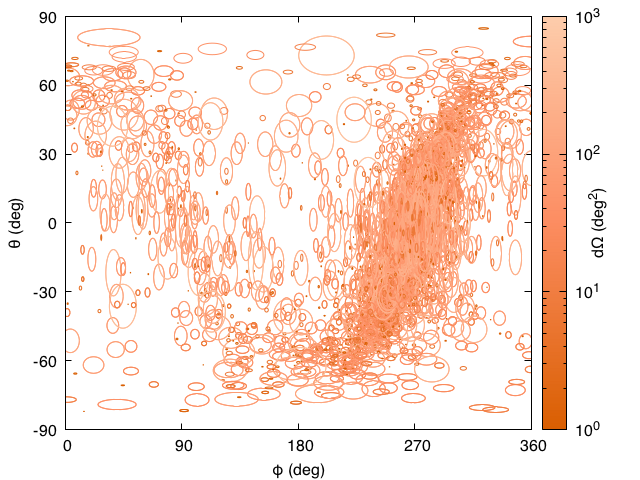} \\
\includegraphics[width=.45\textwidth]{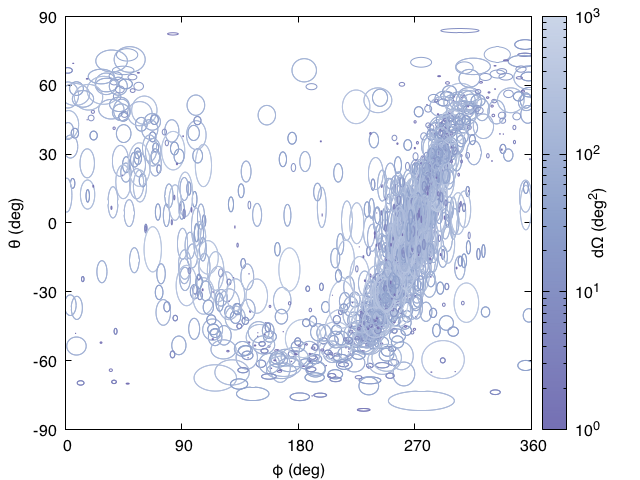} 
\includegraphics[width=.45\textwidth]{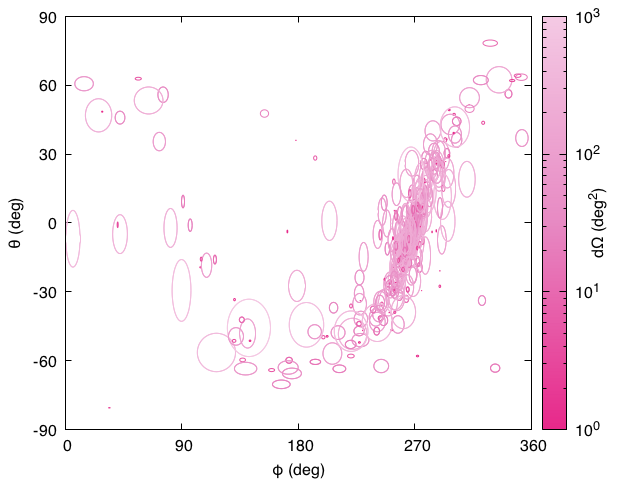} 
    \caption{Sky localisation in ecliptic coordinates for the well-localised binaries. The ellipses encompass the 1$\sigma$ uncertainties on the inferred sky location, and the color scale indicates the angular size of the error region in square degrees.}
    \label{fig:map_LISA}
\end{figure*}

Fig.~\ref{fig:map_LISA} shows the sky localisation and its uncertainty for the different types of binaries. Measuring a sky localisation within 10 square degrees is aided by increased signal to noise, which accumulates throughout the mission for these sources.  Precise sky localisations will typically be available only after 2 years of the mission, or even after 4 years for the fainter systems such the He-He.  The sky maps confirm that~\LISA~will detect DWDs throughout the Galaxy, with important contributions from the bulge and the thick disk. The He-He binaries, which are likely to have the brightest EM counterparts, are overwhelmingly present in the bulge, which will make them very difficult targets due to the large distance and density of the sources.

\subsection{Detection of systems in the stellar halo and beyond}\label{subsec:GW_halo}
Fig.~\ref{fig:hist_freq_LISA} shows that all systems with \fGW $\geqslant$ 3 mHz will be resolved. The latter include systems in the outskirts of the galaxy, including satellite galaxies and streams. \citet{Yu_10_DWD_halo} and \citet{Ruiter_09_halo} found that some DWDs located in the Galactic halo will contribute to the~\LISA~detections. \citet{Korol_18_LISA_LocalGroup} calculated that, given their stellar masses, the Magellanic Clouds and the Andromeda Galaxies are likely to harbour binaries detectable by~\LISA, which may be our only way to constrain the Type Ia supernova rate in our neighbourhood. These studies do not determine whether we will be able to assign these systems to the Galactic halo, the Magellanic Clouds or Andromeda based solely on the GW detections.  

In this section we determine whether~\LISA~will detect any systems belonging to the stellar halo, and whether it will be able to properly identify them as such. Here we define the stellar halo as the ensemble of stars that are present within the virial radius of the main galaxy but were not formed in the main galaxy (ex-situ star formation), although other works sometimes select on present-day distance or kinematics. These stars are typically 8 to 10 Gyrs old.  Our definition of the stellar halo therefore includes satellite galaxies, coherent stellar streams, and the phase mixed remnants of completely disrupted satellites. It does not include stars formed in the main galaxy that were perturbed onto more radial orbits. We directly use the information available from the simulation regarding the position of each star particle at birth to separate in situ (distance from host center < 30 kpc at formation) fromp ex situ (> 30 kpc) stars.

\begin{figure*}
    \centering
    \includegraphics[width=.45\textwidth]{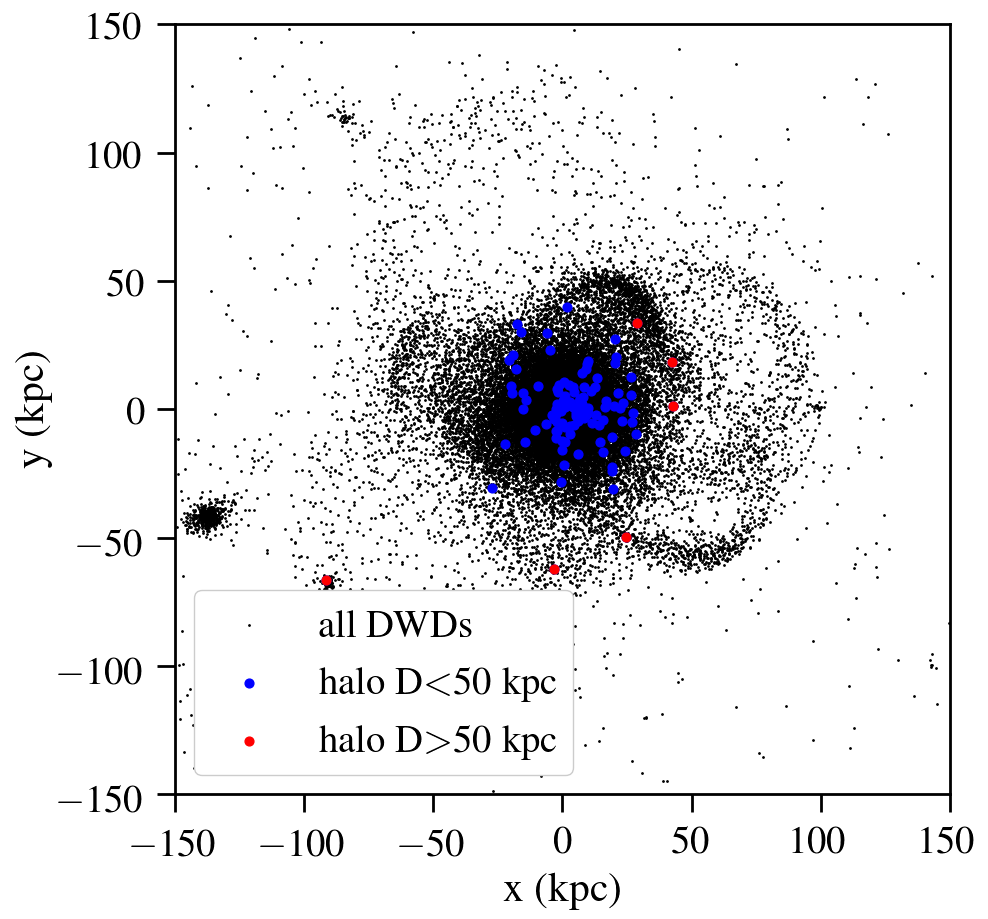}
    \includegraphics[width=.45\textwidth]{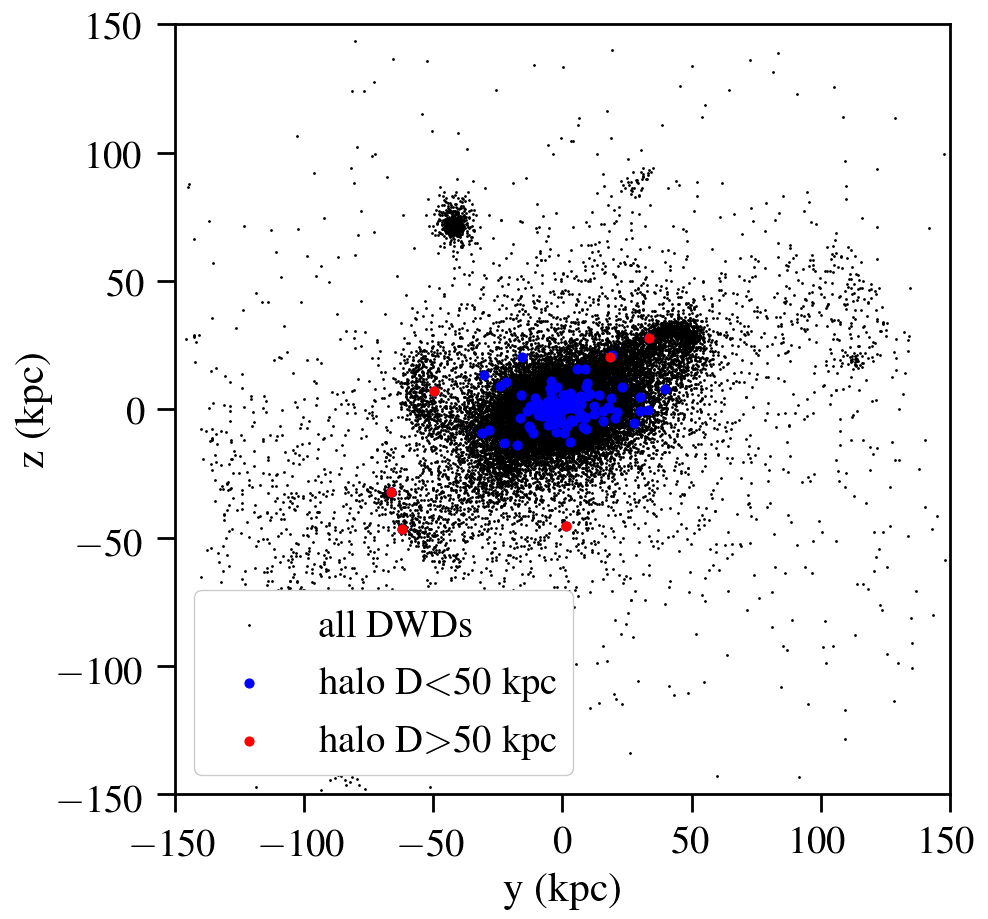}
    \caption{Distribution of  DWDs which have formed outside of the main galaxy (blue dots) and have been accreted. A small fraction of these binaries (red dots) can be attributed to the stellar halo due to their large distance (galactocentric distance larger than 50 kpc), measured with less than 50 per cent accuracy. The complete distribution of DWDs with \fGW$>10^{-4}$ Hz is shown in black, regardless of possible dectectability.}
    \label{fig:halo_systems}
\end{figure*}

\begin{figure}
    \centering
    \includegraphics[width=.45\textwidth]{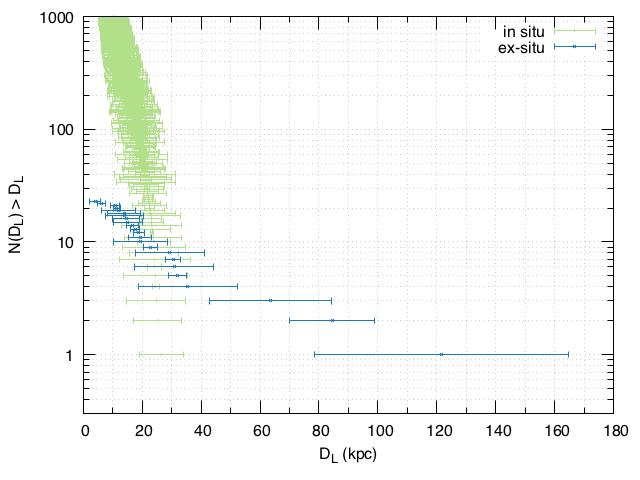}
    \caption{Number of binaries with distance determined to better than 50\%. We distinguish between binaries formed in the main galaxy (in-situ star formation, green) or in satellites (ex-situ star formation, blue). For this simulation (8 year observing time), a handful of the ex-situ binaries have distance measurements which  distinguish them from the other binaries and locate them in satellite galaxies and tidal streams.  }
    \label{fig:cdf_DL_08}
\end{figure}

In total we find that about 5 per cent of the resolved binaries are halo objects. In comparison 1 per cent of the stars in the simulations are halo objects, and DWDs are over-represented in the halo.  Fig.~\ref{fig:halo_systems} shows the localisation of the so-called halo DWDs that have distance measurements with less than 50 per cent uncertainty after an extended mission (8 yrs). 
Many of these DWDs are at small galactocentric distances today such that they overlap spatially with in situ stars (though \citealp{Brown_16_ELM7} found these objects due to their peculiar proper motions).
A few of the halo binaries have measured distances beyond 50 kpc from the center of the Galaxy (red dots) and are effectively located in satellite galaxies and streams. Fig.~\ref{fig:cdf_DL_08} shows the distance of these systems as a function of distance to the Sun.  Only after an extended mission will the uncertainties on the distance measurements become small enough to distinguish the distances of the halo objects from the distance distribution of systems from the main galaxy. These systems typically have sky localisations within 10-30 square degrees, which should be sufficient to assign them to a satellite galaxy.

The number of systems with well measured distances does not depend on the randomly chosen localisation of the Sun, within a ring located at 8.2 kpc from the galactic center. In other words,~\LISA~will produce a complete catalog down to a certain frequency, including the halo systems. However, contrary to the other results in this paper, the number of detected and well-localised halo binaries does depend on the cosmological simulation and its accretion history. Our simulation does not include massive satellite galaxies such as the Magellanic Clouds, which will likely host systems. We can safely speculate that the identification of halo objects around the Milky Way would be possible if the uncertainty of the distance measurement is reduced by more careful selection based on sky position and/or cross-matching with catalogs of satellite galaxies and streams.

\section{Discussion}\label{sec:discussion}

Our model provides the first prediction of DWDs based on a binary population synthesis model combined with a cosmological simulation. The  simulations produce self-consistent star-formation in a cosmological volume.  As such, our work mostly differs from previous models in its assumptions on the galactic structure and star formation history.  Importantly, our model does not rely on  parametrised, axisymmetrical, models of the different components of the MW and self-consistently includes a stellar halo and satellites and streams.

Globally we find that our model predicts comparable numbers of DWDs and of detectable systems than previous studies \citep[e.g.]{nelemans01_WD,Nissanke_2012_DWD} although an exact comparison is impossible due to the different assumptions on the~\LISA~mission (arm length, duration of the mission and S/N threshold for detections) and binary evolution.
Our analysis predicts fewer detectable and well-localised binaries than \cite{Cornish:2017} which sought to update the results of \cite{Nissanke_2012_DWD} with the current \LISA\ design.

\begin{figure}
    \centering
    \includegraphics[width=.45\textwidth]{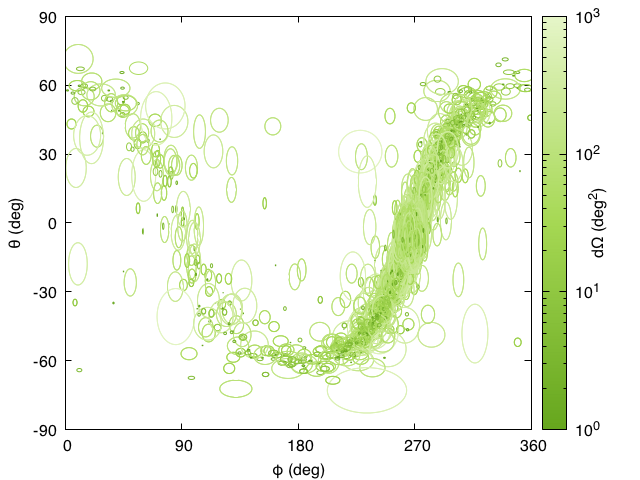}
    \caption{Sky localisation of the individually detected DWDs assuming a spatial distribution following~\citet{nelemans01_WD}, after 8 years of observations. The star formation history and binary evolution model is the same as the rest of the paper. This plot can be directly compared to the sum of the distributions presented in Fig.~\ref{fig:map_LISA}.}
    \label{fig:map_Nelemans}
\end{figure}

The importance of our model lies in its assumptions (or lack thereof) on the galactic structure and typical ages for different stellar populations. Fig.~\ref{fig:map_Nelemans} shows the spatial distribution of the systems detected by~\LISA~ with the Galactic model used in ~\citet{nelemans01_WD}. All other aspects of the model (global metallicity and star formation history and binary evolution) are identical to the model presented here. In comparison with our model shown in Fig.~\ref{fig:map_LISA} this model shows a very thin and well-defined disk, a small bulge and no stellar halo. Quantitatively, the analytic models predict 5 times more systems in the innermost kpc of the Galaxy and about an order of magnitude less beyond 10 kpc. There are almost no detections outside the plane of the disk. Although the localisation of the binaries is different from our model, the global numbers of resolved and well-localised systems is comparable similar to ours. This is probably related to the completeness of the~\LISA~detections down to a a few mHz. Importantly, the simplified model cannot predict the different spatial distributions of the different types of binaries, and its impact on multimessenger astronomy. 

\citet{Ruiter_09_halo} specifically studied the detection of DWDs in the Galactic halo and compute that its signal is a factor 10 lower than the disk's signal. We find somewhat smaller numbers and confirm that He-He systems dominate in the halo. This is because they typically have old stellar progenitors, with a long DWD formation time. The recently formed DWDs have a high enough frequency but have not merged yet. In a detailed study of the thin disk and bulge contributions, \citep{Ruiter_2010_DWD} show no systems with \fGW$>5\times 10^{-2}$ Hz in the bulge, claiming that the latter have all merged in by now. This difference may be related to different assumptions on stellar evolution, or on the simplified model for star formation in the bulge.

\citet{Yu_10_DWD_halo} present an axisymmetric MW model including a thin and thick disk and stellar halo with different characteristic ages for each stellar component. We confirm their distributions of the GW strain and frequency for the different binary types although they find a stronger contribution from He-He systems because of a different treatment for binary interactions. In their model, DWDs undergoing Roche Lobe overflow stay in the~\LISA band and contribute at frequencies \fGW$>5\times10^{-2}$ Hz. They conclude that most of the GW signal comes from the bulge and thin disk and that the thick disk and halo only contribute below 10$^{-3}$ Hz. On the contrary, our  Fig.~\ref{fig:map_LISA} shows that even at high frequencies there is a significant contribution from the thick disk, and a small contribution from the halo in our simulation. We find that the DWDs detected in the halo all have \fGW$>10^{-3}$ Hz, which seems necessary to generate strong signals and be detectable at larger distances. \citep{Yu_10_DWD_halo} explain that all high frequency systems in the halo and disk have merged by now, because of the age of the population. Their different result may also be related to a different binary evolution model. We also note that their sample of DWDs is more than an order of magnitude smaller than ours, and undersampling may lead to the truncation of the binary distribution, especially at high frequency.

Our model is based on a cosmological simulation of a galaxy with strong resemblance to the Milky Way. However, it is not an exact reproduction of the Milky Way. In particular, its present-day star formation is about two times higher than our Milky Way (which has a low present-day star formation rate compared to other galaxies of similar mass) and the scale height of the thin and thick disks in the simulation are about twice as high as observational estimes~\citep{Sanderson2018}. Although the global distribution of satellite galaxies in the simulation is comparable to the Milky Way \citep{Wetzel2016}, our simulation does not contain the equivalent of the Large Magellanic Cloud, which will likely harbour LISA sources.   To understand the limits of our model we performed an identical analysis with the \textbf{m12f} and \textbf{m12m} simulations. In both cases we find that the total number of DWDs, as well as the number of DWDs detectable by~\LISA~ directly scales with the stellar mass of the galaxy. In all simulations the signal is dominated by He-He DWDs, which are over-represented in the bulge, thick disk and stellar halo. They stem from an older, less metal-rich population. There is also a significant contribution from He-CO DWDs, which follow the stellar distribution more closely. In other words, our conclusions regarding different DWD populations are robust. However, quantitative conclusions about the detection of halo objects in the Milky Way would require a model specific model of the Milky Way satellites, which will be possible with current and future EM surveys.

Our model does not include any effects aside from binary interactions. We do not account for triple systems, which have been recently discovered, albeit with wide separations \citep{Perpinya_2018_tripleWD}. We do not  model DWD formation in star clusters, as these low-mass binaries will likely become unbound due to dynamical interactions. We have not modeled the impact of Galactic tides on the binaries, which is negligible for very tight binaries.

The main uncertainty in our model is our choice of binary and stellar evolution parameters. We have chosen "standard" assumptions (such as $\alpha=\lambda=1$ for the common envelope evolution) which are in broad agreement with the current observational constraints and enable comparison with previous work. However, the formation and evolution of the different types of white dwarfs has many uncertainties, influencing the formation time, mass distributions and separation of the binaries. Based on the same cosmological simulations, we plan a more comprehensive study of the parameter space of binary evolution to determine the range of uncertainty we currently have for~\LISA~ detections, both for individual systems and the unresolved background. Similarly, we leave an update of the initial binary properties for future work ~\citep{Moe_2017_binary_props}. Again, one should keeping in mind that by the time~\LISA~operates, many of the uncertainties will have strongly decreased thanks to EM surveys.

\section{Conclusions}\label{sec:conclusions}

We have created the first double white dwarf population model based on the combination of a cosmological simulation of a Milky-Way like galaxy and a binary population synthesis method. We first determine how binaries resulting from the population synthesis map onto the Milky Way galaxy and then how these binaries will be detected by~\LISA. The Milky Way model comes from a cosmological simulation, taken from the FIRE simulation suite, and provides a self-consistent model for star formation which naturally includes all the different galactic components. Their stellar ages and distance distributions lead to distinct contributions to the gravitational wave signal by the different types of white dwarf binaries. In comparison with simplified models, our simulation produces a similar number of detectable sources but we also find many distinct features which are important for the planning of the mission and its scientific exploitation.
\begin{itemize}
    \item Out of the 500 million DWDs in our simulation, over 60 million will be in the~\LISA~frequency band and roughly 12000 will be individually resolved after a nominal mission of 4 years. The catalog will be complete down to a frequency of a few mHz. About 15 per cent of these will be well-enough localised to allow for the search of electromagnetic counterparts with wide-field surveys.
     \item Globally, DWDs follow a similar spatial distribution as stars and are found in all the components of the Galaxy. However, He-He systems, which have formation times up to 12 Gyrs are found among older stellar populations such as the bulge, thick disk and stellar halo. Conversely CO-CO and Ne-X binaries trace young stellar populations and are found in the thin disk. 
     \item High frequency systems (\fGW$>10^{-3}$ Hz), which are the most likely to be well localised by~\LISA~come from recently formed DWDs as they merge quickly, and lower frequency systems need too long timescales to significantly reduce their period. High-frequency systems are dominated by He-He and He-CO systems, which stem from old progenitors and are strongly present in the thick disk and bulge. 
     As such, the sky map of the well-localised systems is very different from previous realisations based on simplified models of the Galaxy.
    \item Above 1 mHz, the Solar neighbourhood is dominated by CO-CO and He-CO binaries, while He-He binaries are 5 times less numerous. The high frequency systems are dominated by He-He and He-CO systems. After 2/4/8 years of~\LISA~60/200/500 systems are expected to be well localised within 5 kpc, which is the maximal distance for detection of regular DWDs with LSST single pointings. With stacked pointings the detection limit can reach close to 10 kpc and 3000 systems could be identified in the 8 year catalog.
      \item With an extended mission, we find that~\LISA~would be able to detect and unambiguously locate systems beyond 50 kpc. The latter are mostly He-He binaries located in satellite galaxies and tidal streams and will have formed outside of the main galaxy. 
  \item Given our binary evolution model, 50 per cent of the resolved systems are He-CO binaries, 30 per cent are He-He binaries, 20 per cent are CO-CO binaries and there are a few per cent of binaries with a Ne WD. Over time, the fraction of He-He systems increases. As~\LISA~will not be able to measure individual component masses, unambiguous classification of the sources will be limited, and model-dependent. The use of additional information may be necessary (e.g. an identified EM counterpart, or the local galactic environment...) 
  
\item Much of the most sought-after information \LISA\ can provide regarding DWDs will become available after long integration times for the observations (e.g., sky localisation) and will benefit from an extended mission (e.g., measurement of the chirp mass). Within the first 2 years, less than 3 per cent of the systems will have a measured chirp mass, which increases up to 8 per cent after 4 years and 15 per cent after 8 years. Towards the end,~\LISA~will measure the lowest chirp masses, which are likely to have the brightest EM counterparts.

    \end{itemize}{}
 These results highlight the importance of refined modeling of complete galactic populations. It is a stepping stone for more thorough analysis of different binary population models, especially in combination with constrains from current and future EM surveys. Our new model also has implications for the preparation of~\LISA~data analysis, including the unresolved gravitational wave background.

\section*{Acknowledgements}
The authors would like to thank Katie Breivik, Silvia Toonen and Valeriya Korol for many exchanges on formation channels of white dwarf binaries and Nelson Christensen and Neil Cornish for many insights on GW detections. Support for AL was provided by the Observatoire de la C\^ote d'Azur and NASA ATP Grant
\#NNX14AH35G, and NSF Collaborative Research Grant
\#1715847 and CAREER grant \#1455342. S.B. is supported by the NSF Graduate Research Fellowship, grant \#DGE1745303. TL acknowledges the support of NASA grant \#NNH15ZDA001N and the NASA LISA Study Office. Support for SGK was provided by NSF Collaborative Research Grant \#1715847 and CAREER grant \#1455342, and NASA grants NNX15AT06G, JPL 1589742, and 17-ATP17-0214. This work was supported by the National Science Foundation through grant PHY 17-148958.

Some of these computations have been done on the Mesocentre SIGAMM machine, hosted by the Observatoire de la C\^ote d'Azur  as well as cluster "Wheeler" hosted by Caltech.

%%%%%%%%%%%%%%%%%%%%%%%%%%%%%%%%%%%%%%%%%%%%%%%%%%

%%%%%%%%%%%%%%%%%%%% REFERENCES %%%%%%%%%%%%%%%%%%

\bibliographystyle{mnras}
\bibliography{dwds,sources}

%%%%%%%%%%%%%%%%%%%%%%%%%%%%%%%%%%%%%%%%%%%%%%%%%%

\bsp	% typesetting comment
\label{lastpage}
\end{document}